\documentclass[aps,prd,twocolumn,preprintnumbers,showpacs,nofootinbib]{revtex4}
\usepackage[dvips]{graphicx}
\usepackage{psfrag}
\usepackage{amsmath}
\advance\oddsidemargin by -\textwidth
\advance\oddsidemargin by -1in
\evensidemargin=\oddsidemargin
\def\fsl#1{\setbox0=\hbox{$#1$}           
   \dimen0=\wd0                                 
   \setbox1=\hbox{/} \dimen1=\wd1               
   \ifdim\dimen0>\dimen1                        
      \rlap{\hbox to \dimen0{\hfil/\hfil}}      
      #1                                        
   \else                                        
      \rlap{\hbox to \dimen1{\hfil$#1$\hfil}}   
      /                                         
   \fi}                                         %

\begin{document}
\preprint{TU-670}
\preprint{KEK-TH-847}

\title{
    Calculating the pion decay constant from $\alpha_s(M_Z)$
}
\author{Michio Hashimoto}
\email[E-mail: ]{michioh@post.kek.jp}
\affiliation{Theory Group, KEK, Oho 1-1, Tsukuba, Ibaraki 305-0801, Japan}
\author{Masaharu Tanabashi}
\email[E-mail: ]{tanabash@tuhep.phys.tohoku.ac.jp}
\affiliation{Department of Physics, Tohoku University, Sendai 980-8578, Japan}

\begin{abstract}
We revisit the analysis of the improved ladder Schwinger-Dyson (SD)
equation for the dynamical chiral symmetry breaking in QCD with
emphasizing the importance of the scale ambiguity.
Previous calculation done so far naively used one-loop 
$\overline{\rm MS}$ coupling in the improved ladder SD equation
without examining the scale ambiguity.
As a result, the calculated pion decay constant $f_\pi$ was less than 
a half of its experimental value $f_\pi=92.4$MeV once the QCD scale is
fixed from the high energy coupling 
$\alpha_s^{\overline{\rm MS}}(M_Z)$.
In order to settle the ambiguity in a proper manner, 
we adopt here in the present paper the next-to-leading-order effective
coupling instead of a 
naive use of the $\overline{\rm MS}$ coupling.
The pion decay constant $f_\pi$ is then calculated from high energy
QCD coupling strength $\alpha_s^{\overline{\rm MS}}(M_Z)=0.1172 \pm
0.0020$.
Within the Higashijima-Miransky
approximation, we obtain
$f_\pi=85$--$106$MeV depending on the value of
$\alpha_s^{\overline{\rm MS}}(M_Z)$ which agrees well with the
experimentally observed value $f_\pi=92.4$MeV\@.
The validity of the improved ladder SD equation is therefore
ascertained more firmly than considered before.
\end{abstract}

\maketitle

\section{Introduction}

The improved ladder approximation of the Schwinger-Dyson (SD)
equation~\cite{Miransky:vj,Miransky:vk}  
has been used widely for the analyses of the dynamical chiral symmetry
breaking in
QCD~\cite{Aoki:1990eq,Jain:1991pk,Jain:1991iy,Kugo:1992zg}
and of models of dynamical electroweak symmetry
breaking~\cite{Yamawaki:1985zg,Miransky:1988xi}. 
In the improved ladder approximation, 
the tree-level one-gluon-exchange diagram, with its QCD coupling
strength being replaced by the running $\overline{\rm MS}$ one, is
assumed to give a dominant contribution in the attractive force
between quark and anti-quark ($q\bar q$).

The method actually succeeded in explaining many quantitative
relations among the low energy hadronic data in
QCD~\cite{Aoki:1990eq,Jain:1991pk,Jain:1991iy,Kugo:1992zg,Aoki:1990yp} 
and established its qualitative 
validity.
Recently, it has also been adopted to predict properties of
QCD under extreme conditions (hot and/or dense 
QCD)~\cite{Barducci:1989wi,Harada:1998zq,Kiriyama:1998na,Hong:1999fh,Roberts:2000aa}.

It is known, however, that a naive use of this approximation, combined
with the high energy QCD coupling such as 
$\alpha_s^{\overline{\rm MS}}(M_Z)$, predicts the pion decay constant
$f_\pi$ significantly smaller than its experimentally observed value 
$f_\pi=92.4$MeV\@.~\cite{Aoki:1990eq,Jain:1991pk}
(See, for example, our Figure~\ref{fig:hig-msbar}, where we obtained
$f_\pi$ of order 30MeV for the one-loop running of the 
$\overline{\rm MS}$ coupling.) 
Note that the decay constant $f_\pi$ determines the scale of chiral
phase transition in QCD\@.
Does this discrepancy indicate the existence of substantial non-ladder
contribution in the driving force of the dynamical chiral symmetry breaking? 
If this is so, 
the success of the improved ladder approximation
should be regarded as an accidental coincidence,
without understanding deeply the nature of the driving force of the
dynamical chiral symmetry breaking.

Failing to explain the experimental value of $f_\pi$ from the high
energy QCD coupling strength,
the improved ladder SD equation is often regarded as a viable
phenomenological model of low energy QCD, in which we can freely
tune its coupling strength so as to make $f_\pi$ consistent with
the experimental value.
We emphasize, however, in order to perform trustful
analysis of QCD under extreme conditions, the low energy models 
need to be connected smoothly with the high energy QCD\@.
The discrepancy of $f_\pi$ thus causes a serious trouble in the
analysis of the critical behavior of hot and/or dense QCD\@.

In this paper, we point out that the previous calculations so far made
for $f_\pi$ overlooked the existence of scale
ambiguity~\cite{Brodsky:1982gc} in the 
improved ladder SD equation. 
In order to settle the ambiguity, 
we introduce a concept of effective 
coupling~\cite{Kennedy:1988sn,Papavassiliou:1991hx} in
the analysis of the improved ladder SD equation of QCD\@.
(The idea similar to our effective coupling was proposed in the
analysis of SD equation by Ref.\cite{Papavassiliou:1991hx} for
different purpose to ours.
It was also adopted by Ref.\cite{Hashimoto:2000uk,Gusynin:2002cu} for
the  analysis of SD equation in gauge theories with extra dimensions.)
The effective coupling is calculated using the background field
method~\cite{Abbott:1981ke} in this paper. 
We next calculate numerically the improved ladder SD equation and
obtain the value of the pion decay constant $f_\pi$ using the high
energy QCD coupling $\alpha_s^{\overline{\rm MS}}(M_Z)$ as an input
parameter of the analysis.
Within Higashijima-Miransky
approximation~\cite{Miransky:vj},  
we obtain $f_\pi=85$--$106$MeV depending on the value of
$\alpha_s^{\overline{\rm MS}}(M_Z)=0.1172 \pm 0.0020$~\cite{Hagiwara:pw}.
The agreement of the calculated $f_\pi$ with the experimentally
observed value $f_\pi=92.4$MeV (or the value in the chiral limit
$m_u=m_d=m_s=0$, $f_\pi=86$MeV~\cite{Gasser:1984gg}) is rather impressing.
The discrepancy of $f_\pi$ is resolved in our analysis.
We thus ascertained the validity of the improved ladder SD equation
more firmly than considered before.


We organize the present paper as follows:
In Section~\ref{sec:ladder-sd} we give a brief derivation of the SD
equation and point out the problem of the scale ambiguity.
In Section~\ref{sec:effective} the concept of effective coupling is
introduced. 
The behavior of QCD coupling and its regularization are discussed in
Section~\ref{sec:running}.
In Section~\ref{sec:higashijima} we give our results of numerical
analysis within the Higashijima-Miransky approximation to the angular
integral of the SD equation.
A numerical analysis without the Higashijima-Miransky approximation is
performed in Section~\ref{sec:non-local} using the non-local gauge
fixing parameter method.
Section~\ref{sec:summary} is devoted for summary and discussions.

\section{Ladder Schwinger-Dyson equation }
\label{sec:ladder-sd}

We outline here a derivation of the
ladder SD equation based on
Ref.\cite{Kleinert:1976xz,Shrauner:mc,Kugo:1978ct}.
We first integrate out the gluon field from the QCD Lagrangian at the tree
level.
We then obtain a bi-local interaction model,
\begin{eqnarray}
\lefteqn{
  S_{\rm BL} =
  \int \!\! d^4 x \, \bar\psi i\fsl{\partial} \psi
  + \frac{1}{2} \int d^4 x_1 d^4 x_2 \times 
} \nonumber\\
  & & \times (\bar \psi \gamma^\mu T^a \psi)_{x_1}
    (\bar \psi \gamma^\nu T^a \psi)_{x_2} \tilde D_{\mu\nu}(x_1-x_2),
  \label{eq:bi-local}
\end{eqnarray}
with gluon propagator $\tilde D_{\mu\nu}(x)$ being given by
\begin{eqnarray}
\lefteqn{  
  \tilde D_{\mu\nu}(x) \equiv
} \nonumber\\
  & &  \int \!\! \dfrac{d^4 k}{(2\pi)^4}
    e^{-ik\cdot x}
    \dfrac{g_s^2}{k^2}\left(
      g_{\mu\nu} - (1-\xi)\dfrac{k_\mu k_\nu}{k^2}
    \right).
\end{eqnarray}
Here $g_s$ and $\xi$ are the QCD coupling constant and the gauge
parameter, respectively.
In Eq.(\ref{eq:bi-local}) we have neglected six or higher quark
interaction terms.
The ladder SD equation for the dynamical chiral symmetry breaking
can be
derived from Eq.(\ref{eq:bi-local}) (bi-local four fermion
interaction) in a similar manner to the gap equation in the NJL model 
(local four fermion interaction).
We obtain
\begin{subequations}
\label{eq:sdS}
\begin{eqnarray}
\lefteqn{
  A(-p^2) = 
    1 + \dfrac{C_F}{-p^2}\int \!\! \dfrac{d^4 q}{(2\pi)^4 i} \times
} \nonumber\\
    & & \times
        \dfrac{A(-q^2)}{-A^2(-q^2) q^2 + B^2(-q^2)} 
    g_s^2 \left[
      (1+\xi) \dfrac{p\cdot q}{(p-q)^2} 
    \right.
    \nonumber\\
    & & \quad\quad \left.
     +2(1-\xi)\dfrac{p\cdot (p-q) q \cdot (p-q)}{(p-q)^4}
    \right], 
\label{eq:sdA} \\
\lefteqn{
  B(-p^2) = C_F \int \!\! \dfrac{d^4 q}{(2\pi)^4 i} \times
} \nonumber\\
  & &  \times
        \dfrac{B(-q^2)}{-A^2(-q^2) q^2 + B^2(-q^2)}
        g_s^2 \dfrac{3+\xi}{-(p-q)^2},
\label{eq:sdB} 
\end{eqnarray}
\end{subequations}
with $C_F \equiv T_a T_a = (N_c^2-1)/(2N_c)=4/3$ being the Casimir of
the fundamental representation of $SU(N_c=3)$.
The functions $A(-p^2)$ and $B(-p^2)$ in Eq.(\ref{eq:sdS}) are quark
wave-function and mass-function, respectively.
The quark propagator $S$ is given in terms of these functions,
\begin{equation}
  iS^{-1}(p)=\fsl{p} A(-p^2) - B(-p^2).
\end{equation}
Nonvanishing mass-function $B \ne 0$ implies the dynamical chiral
symmetry breaking in QCD\@.
It is therefore regarded as an order parameter of this system.

Performing the Wick rotation and the angular integrals, we obtain
\begin{subequations}
\label{eq:sdS2}
\begin{eqnarray}
  A(x) &=& 1+\dfrac{C_F}{x} \int_0^{\Lambda^2} \!\!\!\!\!\! dy 
     \dfrac{yA(y)}{yA^2(y)+B^2(y)}K_A(x,y),
  \nonumber\\
  & &\\
  B(x) &=& C_F \int_0^{\Lambda^2} \!\!\!\!\!\! dy 
     \dfrac{yB(y)}{yA^2(y)+B^2(y)}K_B(x,y).
\end{eqnarray}
\end{subequations}
Here 
we have introduced the ultraviolet (UV) cutoff $\Lambda$ in the SD
equation. 
It is known that the result of the SD equation is insensitive to the
cutoff $\Lambda$ if we take sufficiently large $\Lambda$ in QCD\@.
The integral kernels $K_A$ and $K_B$ of the SD equation
Eq.(\ref{eq:sdS2}) are given by
\begin{subequations}
\label{eq:kernel}
\begin{eqnarray}
  K_A(x,y) &\equiv& \dfrac{1}{2\pi^2}\int_0^\pi \!\! 
  d\theta \sin^2\theta \times
  \nonumber\\
  & & \times 
  \alpha_s\Bigl[
    (3-\xi)\dfrac{\sqrt{xy}\cos\theta}{z}
  \nonumber\\
  & & 
   \quad -2(1-\xi)\dfrac{xy}{z^2}\sin^2\theta
  \Bigr], 
  \\
  K_B(x,y) &\equiv& \dfrac{1}{2\pi^2}\int_0^\pi\!\! d\theta  
  \sin^2\theta 
  \dfrac{(3+\xi)\alpha_s}{z}, 
\end{eqnarray}
\end{subequations}
with $z$ and $\alpha_s$ being the square of the gluon momentum,
\begin{equation}
  z\equiv x+y-2\sqrt{xy}\cos\theta
\end{equation}
and the QCD coupling
\begin{equation}
  \alpha_s \equiv \dfrac{g_s^2}{4\pi},
\end{equation}
respectively.

The ladder SD equation Eq.(\ref{eq:sdS2}) combined with the kernel
Eq.(\ref{eq:kernel}) does not take account of the running of QCD
coupling strength, however.
In order to make the ladder SD equation consistent with the
renormalization group (RG) consideration, the bi-local four-fermion
interaction in Eq.(\ref{eq:bi-local}) needs to be modified to include
its running effects. 
A widely adopted prescription for such a purpose is to
replace the QCD coupling $g_s$ with the $\overline{\rm MS}$
renormalized one $g_s(\mu)$ where $\mu$ is identified with the gluon
momentum
$\sqrt{z}$.~\cite{Miransky:vj,Lane:he,Sugawara:1984vw,Roberts:mj}

The prescription is not unique, however.
Actually, we can use $g_s^2(\mu=c\sqrt{z})$ ($c\ne 1$, $c>0$)
instead of the conventional choice $g_s^2(\mu=\sqrt{z})$ (scale
ambiguity~\cite{Brodsky:1982gc}).
Even if we adopt $c\ne 1$, the solution of the SD equation is shown to 
be consistent with the RG\@.
Moreover, as we stated previously, naive analysis based on the
improved ladder SD equation with $c=1$ predicts $f_\pi$ much lower
than its experimentally observed value
$f_\pi=92.4$MeV\@.~\cite{Aoki:1990eq,Jain:1991pk}

In order to calculate the pion decay constant $f_\pi$ from the high
energy QCD coupling $\alpha_s(M_Z)$, we thus need to resolve the scale 
ambiguity in the context of the improved ladder SD equation.
We consider this problem in the next section by introducing the
concept of effective coupling.

\section{Effective coupling in QCD}
\label{sec:effective}

Before investigating the scale ambiguity of the improved ladder SD
equation of QCD, we first consider simpler model, the strongly coupled 
QED with $N$ flavor of massless
fermions~\cite{Kondo:1988md,Gusynin:1989mc}.

In this model, ladder SD equation can be derived from a bi-local
four-fermion interaction
\begin{equation}
  (\bar\psi\gamma^\mu\psi)_{x_1}(\bar\psi\gamma^\mu\psi)_{x_2}
  \tilde D_{\mu\nu}^{\rm QED}(x_1-x_2),
\label{eq:QED-bilocal}
\end{equation}
which is induced by the photon propagator
\begin{eqnarray}
\lefteqn{  
  \tilde D^{\rm QED}_{\mu\nu}(x) \equiv
} \nonumber\\
  & &  \int \!\! \dfrac{d^4 k}{(2\pi)^4}
    e^{-ik\cdot x}
    \dfrac{g_e^2}{k^2}\left(
      g_{\mu\nu} +\mbox{(gauge fixing)}
    \right),
\nonumber\\
  & &
\label{eq:photon_prop}
\end{eqnarray}
with $g_e$ being the QED coupling constant.

Thanks to the Ward-Takahashi (WT) identities of QED, 
it is enough to calculate vacuum polarization function (loop
correction to the photon propagator) for the evaluation of the
bi-local four-fermion interaction Eq.(\ref{eq:QED-bilocal}).
At the one-loop level, the $g_e^2/k^2$ part of
Eq.(\ref{eq:photon_prop}) is then replaced as
\begin{eqnarray}
  \dfrac{g_e^2}{k^2} &\rightarrow&
  \dfrac{g_e^2}{\left(1-g_e^2 \Pi^{\rm QED}(k^2)\right)k^2}
  \nonumber\\
  &=& 
  \dfrac{1}{\left( \dfrac{1}{g_{e{\overline{\rm MS}}}^2(\mu)}
               - \Pi^{\rm QED}_{\overline{\rm MS}}(k^2,\mu)
            \right)k^2 },
\label{eq:eff_QED1}
\end{eqnarray}
with $g_{e{\overline{\rm MS}}}$ being the $\overline{\rm MS}$
renormalized QED coupling.
The $\overline{\rm MS}$ renormalization is performed in the second
line of Eq.(\ref{eq:eff_QED1}).
It is straightforward to calculate the $\overline{\rm MS}$ regularized
vacuum polarization function $\Pi^{\rm QED}_{\overline{\rm MS}}$:
\begin{equation}
  \Pi^{\rm QED}_{\overline{\rm MS}}(k^2,\mu) 
  = \frac{4}{3} N \left[ 
      \ln\dfrac{-k^2}{\mu^2} - \frac{5}{3}
    \right].
\end{equation}
We compare Eq.(\ref{eq:eff_QED1}) with the prescription of the naive
improved ladder approximation, 
\begin{equation}
  \dfrac{g_e^2}{k^2} \rightarrow
  \dfrac{g_{e{\overline{\rm MS}}}^2(\mu=\sqrt{|k^2|})}{k^2}.
\label{eq:naive_imp}
\end{equation}
The naively improved formula Eq.(\ref{eq:naive_imp}) differs from our
one-loop formula Eq.(\ref{eq:eff_QED1}) due to the existence of
nonvanishing finite part in the vacuum polarization,
\begin{equation}
  \Pi^{\rm QED}_{\overline{\rm MS}}(k^2,\mu = \sqrt{|k^2|}) 
  = -\frac{20}{9} N, \quad
  -k^2 \ge 0.
\end{equation}

One method to resolve the scale ambiguity is to use the
renormalization scale $\mu$ so as to minimize the finite corrections. 
In the present model, we find that the choice 
\begin{equation}
  \mu = \sqrt{|k^2|} \exp\left(-\frac{5}{6}\right)
\label{eq:QED_scale}
\end{equation}
eliminates the finite part
\begin{equation}
  \Pi^{\rm QED}_{\overline{\rm MS}}(k^2,
  \mu = \sqrt{|k^2|}\exp\left(-\frac{5}{6}\right)) = 0.
\end{equation}

The other method is to use the effective coupling in the improved
ladder SD equation.
The effective coupling is defined by
\begin{equation}
  \dfrac{1}{g_{\rm eff}^2(-k^2, \mu)}
  \equiv
  \dfrac{1}{g_{e{\overline{\rm MS}}}^2(\mu)} 
               - \Pi^{\rm QED}_{\overline{\rm MS}}(k^2,\mu).
\label{eq:eff_QED2}
\end{equation}
The $\mu$-dependence in the RHS of Eq.(\ref{eq:eff_QED2}) cancels
at the leading-order.
This method has an advantage compared with the former one that it
can easily deal with massive particles in loop.
We therefore adopt the effective coupling method hereafter.

We now turn to the problem of the scale ambiguity of the improved
ladder SD equation of QCD\@.
As we have done in the case of QED, we need to calculate loop
corrections to the bi-local four-fermion interaction
Eq.(\ref{eq:bi-local}) in QCD\@.
A difficulty arises, however, in the case of QCD, where Slavnov-Taylor 
(ST) identities hold instead of the QED-like WT identities.
Due to the lack of the QED-like WT identities, 
evaluation of the vacuum polarization function is not enough to
renormalize the QCD coupling constant.

\begin{figure}[htbp]
  \begin{center}
    \includegraphics[width=4cm]{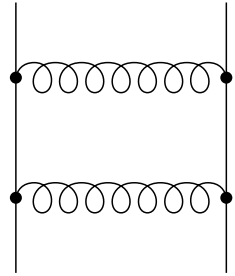}
  \end{center}
  \caption{Ladder-like diagram which is included in the definition of
      the PT effective coupling.}
  \label{fig:ladder-like-graph}
\end{figure}

Pinch technique (PT) method developed by Cornwall and
Papavassiliou~\cite{Cornwall:1981zr,Cornwall:1989gv}
may be a hopeful candidate to resolve the scale ambiguity in the
improved ladder SD equation of QCD\@.
The effective coupling in the (S-matrix) PT is extracted from the
on-shell  $q\bar q$ scattering amplitude.
Since the renormalization scale dependence should cancel in the
on-shell amplitude, the scale ambiguity in the $q\bar q$ scattering
amplitude is automatically resolved in the PT\@.
In addition, it does not depend on the choice of the gauge parameter.
We note, however, that these great features of the PT effective
coupling is assured only when contributions of the pinch part of the
ladder type diagram 
Figure~\ref{fig:ladder-like-graph} is included in the amplitude.
The pinch part of Figure~\ref{fig:ladder-like-graph} gives non-zero
amplitude in gauges other than the Feynman gauge $\xi=1$.
The ladder type diagrams are resummed to all orders in the ladder SD
equation.
So, some portion of the ladder contribution will be doubly counted if
we simply adopt the PT effective coupling in the analysis of the improved
ladder SD equation with $\xi\ne 1$.

We therefore adopt here a slightly different choice, the background
gauge fixing method~\cite{Abbott:1981ke}, where the QED-like WT
identities hold even in the case of QCD\@.
Thanks to the naive QED-like WT identities, the QCD coupling can be
easily renormalized by the calculation of the vacuum polarization
function in this method
\begin{equation}
  \dfrac{1}{g_{\rm eff}^2(-k^2,\mu)}
  \equiv
    \dfrac{1}{g_{s\overline{\rm MS}}^2(\mu)}
   - \Pi_{\overline{\rm MS}}^{\rm QCD}(k^2, \mu),
\label{eq:geff0}
\end{equation}
with $\Pi_{\overline{\rm MS}}^{\rm  QCD}(k^2,\mu)$
being the $\overline{\rm MS}$ regularized vacuum polarization
function.
The $\overline{\rm MS}$ renormalized QCD coupling is denoted by
$g_{s\overline{\rm MS}}$.
Using the one-loop finite part of 
$\Pi_{\overline{\rm MS}}^{\rm  QCD}(k^2,\mu)$, it is easy to show that 
the $\mu$-dependence cancels in the Eq.(\ref{eq:geff0}) at the
leading-order. 
We note that the coupling Eq.(\ref{eq:geff0}) can be free from the
double 
counting problem in the ladder SD equation, since it does not include
contributions from the ladder type diagrams.

After a straightforward calculation, we obtain
\begin{eqnarray}
  \Pi^{\rm QCD}_{\overline{\rm MS}}(k^2,\mu)
  &=& 
 C_G \biggl[ 
    4 I_0^R(k^2,\mu) - I_1^R(k^2,\mu) 
 \nonumber\\
 & &
   - 2\dfrac{(1-\xi_{\rm bg})}{(4\pi)^2}
   + \frac{1}{4}\dfrac{(1-\xi_{\rm bg})^2}{(4\pi)^2}
  \biggr]
 \nonumber\\
 & &  - 8T_R \sum_{f=1}^{N_F} I_2^R(k^2, m_f^2,\mu).
\label{eq:QCD_finite}
\end{eqnarray}
with $C_G=N_c=3$, $T_R = 1/2$ for QCD ($SU(N_c=3)$ gauge theory) and
$\xi_{\rm bg}$ being the gauge fixing parameter in the background
gauge.
The number of flavors is denoted by $N_F$.
Here the functions $I_0^R$, $I_1^R$ are given by
\begin{subequations}
\begin{eqnarray}
  (4\pi)^2 I_0^R(k^2,\mu)
  &=& -\ln\dfrac{-k^2}{\mu^2} + 2, 
\label{eq:I0R} \\
  (4\pi)^2 I_1^R(k^2,\mu)
  &=& -\frac{1}{3}\ln\dfrac{-k^2}{\mu^2} + \frac{5}{9}.
\end{eqnarray}
\end{subequations}
The function $I_2^R$ represents the quark loop contribution.
We obtain
\begin{eqnarray}
\lefteqn{
  (4\pi)^2 I_2^R(k^2, m_f^2) = -\frac{1}{6}\ln\dfrac{m_f^2}{\mu^2}
} \nonumber\\
  & & +\frac{1}{6} \left\{
         \dfrac{1}{X_f^2}\left[
           \dfrac{1}{X_f} \tanh^{-1} X_f - 1 - \frac{1}{3} X_f^2
         \right]
       \right.
       \nonumber\\
  & & \qquad\qquad
        \left. -3\left[
           \dfrac{1}{X_f} \tanh^{-1} X_f - 1
         \right]
       \right\},
\label{eq:I2R}
\end{eqnarray}
with
\begin{equation}
  X_f \equiv \sqrt{\dfrac{-k^2}{4m_f^2 - k^2}}.
\end{equation}
For a massless quark $m_f=0$, the expression Eq.(\ref{eq:I2R}) reads 
\begin{equation}
  (4\pi)^2 I_2^R(k^2,m_f^2=0)
  = -\frac{1}{6}\ln\dfrac{-k^2}{\mu^2} + \frac{5}{18}.
\label{eq:I2R0}
\end{equation}
Note here that the Eq.(\ref{eq:geff0}) depends on the choice of the
gauge fixing parameter $\xi_{\rm bg}$.
It is also known that the effective coupling Eq.(\ref{eq:geff0}) with
$\xi_{\rm bg}=1$ coincides with the PT effective
coupling.~\cite{Denner:1994nn,Binosi:2002ft}  

Such a $\xi_{\rm bg}$-dependence cancels with the ladder-type
contributions and other vertex corrections in the one-loop amplitude.
The ladder-type diagrams are resummed in the analysis of the ladder SD
equation with a particular gauge parameter $\xi$.
It is therefore plausible to take $\xi_{\rm bg}=\xi$.
As we will show in section~\ref{sec:higashijima}, the most convenient
gauge parameter is $\xi=0$ in the analysis of the ladder SD equation
with Higashijima-Miransky approximation. 
We will therefore take $\xi_{\rm bg}=0$ in the
analysis of following sections.
We will also use $\xi_{\rm bg}=1$ (the PT effective coupling) in order
to make a comparison between them.

We comment on the choice of the renormalization scale $\mu$ which
minimizes the finite correction Eq.(\ref{eq:QCD_finite}).
It is easy to find, for $N_F=0$, 
the finite correction vanishes with the renormalization
scale $\mu=\sqrt{|k^2|}\exp(-205/264)$ for $\xi_{\rm bg}=0$
($\mu=\sqrt{|k^2|}\exp(-67/66)$ for $\xi_{\rm bg}=1$).

\section{Behavior of running coupling}
\label{sec:running}

We next evaluate numerically the behavior of the running couplings of
QCD\@. 
Since we are interested in the dynamical chiral symmetry breaking, we
consider the running below the scale of b-quark mass $m_b=4.3$GeV\@.

In our analysis, we use the PDG average~\cite{Hagiwara:pw} of the high
energy QCD coupling $\alpha_s^{\overline{\rm MS}}(\mu=M_Z)$,
\begin{equation}
  \alpha_s^{\overline{\rm MS}}(M_Z) = 0.1172 \pm 0.0020,
  \quad
  N_F=5,
\label{eq:alphas_mz}
\end{equation}
where the $\overline{\rm MS}$ coupling is defined in QCD with $N_F=5$
flavor of quarks.
Eq.(\ref{eq:alphas_mz}) corresponds to 
\begin{equation}
  \alpha_s^{\overline{\rm MS}}(m_b) = 0.2197 \pm 0.0075, \quad
  N_F=4,
\label{eq:bc_rge}
\end{equation}
at the next-to-next-to-leading-order (NNLO), and
\begin{equation}
  \alpha_s^{\overline{\rm MS}}(m_b) = 0.2188 \pm 0.0074, \quad
  N_F=4,
\label{eq:bc_rge2}
\end{equation}
at the next-to-leading-order (NLO) approximation.
We note that the difference between NLO and NNLO is of negligible
order at $\mu=m_b$.

\begin{figure}[htbp]
  \begin{center}
    \psfrag{alphas}[][]{{\LARGE $\alpha_s$}}
    \psfrag{mu/q2 [GeV]}[][]{\large {$\mu$ or $\sqrt{-k^2}$ [GeV]}}
    \psfrag{1loop MSb}[Bl][Bl]{\footnotesize 
      one-loop $\overline{\rm MS}$}
    \psfrag{2loop MSb}[Bl][Bl]{\footnotesize
      two-loop $\overline{\rm MS}$}
    \psfrag{Landau}[Bl][Bl]{\footnotesize
      $\xi_{\rm bg}=0$}
    \psfrag{Feynman}[Bl][Bl]{\footnotesize
      $\xi_{\rm bg}=1$}
    \includegraphics[width=6.5cm]{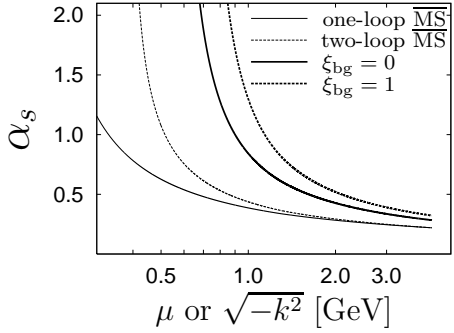}
  \end{center}
\caption{Running of the $\overline{\rm MS}$ (one- and two-loop) 
  and effective ($\xi_{\rm bg}=0$ and $\xi_{\rm bg}=1$) couplings 
  in QCD\@.
  The boundary condition of RGE is taken to be 
  $\alpha_s^{\overline{\rm MS}}(m_b)=0.2197$, which  corresponds to
  $\alpha_s^{\overline{\rm MS}}(M_Z)=0.1172$ at the scale of $M_Z$.
}
\label{fig:running}
\end{figure}

The renormalization group equation (RGE) of $\overline{\rm MS}$
coupling is solved numerically below the scale of $m_b$ using
Eq.(\ref{eq:bc_rge}) as its boundary condition. 
As we noted before, the difference between Eq.(\ref{eq:bc_rge}) and
Eq.(\ref{eq:bc_rge2}) is insignificant.
We assume $m_u = m_d = m_s = 0$ for three light quarks.
The $N_F=4$ $\overline{\rm MS}$ RGE is adopted for 
$\mu\ge m_c=1.3$GeV, while $N_F=3$ is used for $\mu \le m_c$.
Figure~\ref{fig:running} shows the running behavior of one- and
two-loop $\overline{\rm MS}$ couplings.
Note here that the two-loop effects become important for 
$\mu\lesssim 1$GeV\@.

We emphasize here that we do not use
Eq.(\ref{eq:alphas_mz}) directly as a boundary condition of the
one-loop RGE\@.
Actually, the two-loop RGE effect between $M_Z$ and $m_b$ is of
non-negligible order.
If we adopt directly Eq.(\ref{eq:alphas_mz}) as the RGE boundary
condition, the one-loop RGE leads to the value of
$\alpha_s^{\overline{\rm MS}}(m_b)$ sizably smaller than the
values of Eq.(\ref{eq:bc_rge}) and Eq.(\ref{eq:bc_rge2}).
The behavior of one-loop $\overline{\rm MS}$ coupling with this
boundary condition is then quite different from the result of
Figure~\ref{fig:running}. 

We next investigate the behavior of the effective coupling.
By using Eqs.(\ref{eq:geff0}--\ref{eq:I2R},\ref{eq:I2R0}) it is easy
to show
\begin{eqnarray}
\lefteqn{
  \dfrac{4\pi}{\alpha_s^{\rm eff}(-k^2, \mu \ge m_c)} = 
} \nonumber\\
  & & 
  \dfrac{4\pi}{\alpha_s^{\overline{\rm MS}}(\mu)}
   -\Pi_{\rm light} - \Pi_{\rm charm} 
   -\frac{4}{3} T_R \ln\dfrac{m_c^2}{\mu^2}.
  \nonumber\\
  & &
\label{eq:geffhigh}
\end{eqnarray}
Here $\Pi_{\rm light}$ and $\Pi_{\rm charm}$ are defined as
\begin{subequations}
  \begin{eqnarray}
    \Pi_{\rm light} &\equiv& 
       -C_G \left[
         \frac{11}{3}\ln\dfrac{-k^2}{\mu^2} - \frac{67}{9}
       \right]
       \nonumber\\
       & & - C_G \left[
          2(1-\xi_{\rm bg}) - \frac{1}{4} (1-\xi_{\rm bg})^2 
       \right]
       \nonumber\\
       & & +\frac{4}{3} T_R N_{\ell} \left[
       \ln\dfrac{-k^2}{\mu^2} - \frac{5}{3}
       \right], 
     \\
     \Pi_{\rm charm} &\equiv&
     \frac{4}{3} T_R 
       \left\{
         3\left[
           \dfrac{1}{X_c} \tanh^{-1} X_c - 1
         \right]
        \right.
      \nonumber\\
      & & \left.
        -\dfrac{1}{X_c^2}\left[
           \dfrac{1}{X_c} \tanh^{-1} X_c - 1 - \frac{1}{3} X_c^2
         \right]
       \right\},
      \nonumber\\
      & &
  \end{eqnarray}
\end{subequations}
with $N_\ell$ being the number of light flavors $N_\ell=3$.
The parameter $X_c$ is defined as
\begin{equation}
  X_c \equiv \sqrt{\dfrac{-k^2}{4m_c^2 - k^2}}.
\end{equation}

The $\overline{\rm MS}$ coupling $\alpha_s^{\overline{\rm MS}}(\mu)$ in
Eq.(\ref{eq:geffhigh}) is the coupling at the scale $\mu\ge m_c$ and
thus defined in 4 flavor QCD\@.
For $\alpha_s^{\overline{\rm MS}}(\mu<m_c)$ (3 flavor QCD), the 
$\ln m_c^2/\mu^2$ term is just missing, 
\begin{eqnarray}
\lefteqn{
  \dfrac{4\pi}{\alpha_s^{\rm eff}(-k^2, \mu<m_c)} = 
} \nonumber\\
  & &  
\dfrac{4\pi}{\alpha_s^{\overline{\rm MS}}(\mu)}
   -\Pi_{\rm light} - \Pi_{\rm charm}.
\label{eq:geff}
\end{eqnarray}
By using the expansion of $\tanh^{-1} X_c$ around $X_c=0$,
\begin{eqnarray}
  \tanh^{-1} X_c 
  &=& \frac{1}{2}\ln\left(\dfrac{1+X_c}{1-X_c}\right)
  \nonumber\\
  &=& X_c + \frac{1}{3} X_c^3  +  \frac{1}{5} X_c^5 + \cdots,
\end{eqnarray}
it is easy to check the decoupling of the c-quark effect in the 
$m_c \rightarrow \infty$ limit in Eq.(\ref{eq:geff})

We show the behavior of the effective coupling for $\xi_{\rm bg}=0$
and $\xi_{\rm bg}=1$ in Figure~\ref{fig:running}.
Since we want to calculate them at NLO level,
we adopt the two-loop $\overline{\rm MS}$ coupling in
Eq.(\ref{eq:geffhigh}) and Eq.(\ref{eq:geff}).
We also assumed $\mu=\sqrt{-k^2}$ in the plot.
We note that the effective coupling $\alpha_s^{\rm eff}$ is
significantly larger than the $\overline{\rm MS}$ one.

\begin{figure}[htbp]
  \begin{center}
    \psfrag{alphas}[][]{{\LARGE $\alpha_s$}}
    \psfrag{mu/q2 [GeV]}[][]{\large {$\sqrt{z}$ [GeV]}}
    \psfrag{al0}[][]{\large {$\alpha_0$}}
    \psfrag{al1}[][]{\large {$\alpha_1$}}
    \includegraphics[width=6.5cm]{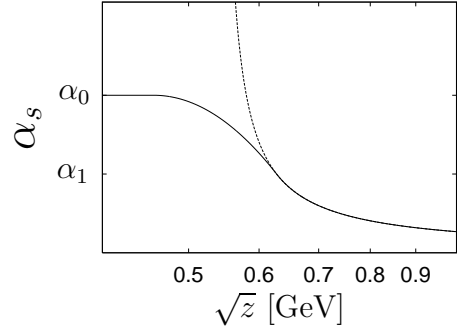}
  \end{center}
\caption{Typical behavior of the IR regularized effective coupling.
}
\label{fig:IR-regularization}
\end{figure}

Both the $\overline{\rm MS}$ coupling 
$\alpha_s^{\overline{\rm MS}}(\mu=\sqrt{-k^2}=\sqrt{z})$ 
and the effective coupling 
$\alpha_s^{\rm eff}(-k^2=z, \mu=\sqrt{-k^2}=\sqrt{z})$ 
diverge in the infrared (IR) region.
We thus need to regularize them in order to solve the improved ladder SD
equation with these coupling strengths.
The IR regularized coupling $\alpha_s(z)$ is regularized such that 
$d\alpha_s(z)/dz$ is continuous for $\Lambda^2 \ge z \ge 0$.
($\Lambda$ is the UV cutoff of the SD equation.)
In this paper, we adopt an IR regularization scheme~\cite{Aoki:1990eq},
\begin{eqnarray}
\lefteqn{
  \alpha_s(z = \Lambda^2 e^t) 
} \nonumber\\
  & & 
  =
  \begin{cases}
    \alpha_0
      & \mbox{for $t\le t_0$}, \\
    \alpha_0 - \frac{1}{2} C(t-t_0)^2
      & \mbox{for $t_0 \le t \le t_1$}, \\
    \alpha_s^{\rm eff}(z=\Lambda^2 e^t)
      & \mbox{for $t_1 \le t$},
  \end{cases}
\label{eq:IR-regularization}
\end{eqnarray}
with $t_1$, $t_0$, $C$ being determined by
\begin{subequations}
\begin{eqnarray}
  & & \alpha_s^{\rm eff}(z=\Lambda^2 e^{t_1}) = \alpha_1, \\
  & & t_0 = t_1 + \dfrac{2(\alpha_0-\alpha_1)}{
      \left. z\dfrac{d}{dz} 
      \alpha_s^{\rm eff} \right|_{z=\Lambda^2 e^{t_1}}},
  \\
  & & C = \dfrac{1}{2(\alpha_0 - \alpha_1)}
          \left[\left.
            z\frac{d}{dz} \alpha_s^{\rm eff} \right|_{z=\Lambda^2 e^{t_1}}
          \right]^2.
\end{eqnarray}
\end{subequations}
We also adopt the same IR regularization for the $\overline{\rm MS}$
coupling. 
The typical behavior of the IR regularized coupling $\alpha_s$ is
depicted in Figure~\ref{fig:IR-regularization}.
Note here that our IR regularization Eq.(\ref{eq:IR-regularization}) 
contains two IR coupling parameters $\alpha_0$ and $\alpha_1$.
We regard the result of the SD equation reliable only when they
are not too sensitive to these IR parameters.

\section{Calculating $f_\pi$ with Higashijima-Miransky approximation}
\label{sec:higashijima}

The angular integral Eq.(\ref{eq:kernel}) cannot be performed in an
analytical manner.
In this section, we consider so called
Higashijima-Miransky~\cite{Miransky:vj}
approximation,
in which $\alpha_s(z)$ in Eq.(\ref{eq:kernel}) is replaced as
\begin{equation}
  \alpha_s(z) \rightarrow \alpha_s(\mbox{max}(x,y)).
\label{eq:higashijima-type}
\end{equation}
We can then analytically perform the angular integral.
We obtain
\begin{subequations}
  \begin{eqnarray}
    K_A(x,y) &=& \dfrac{\xi}{4\pi}\left[
      \alpha_s(x) \frac{y}{x}\theta(x-y) + (x\leftrightarrow y)
    \right], 
  \nonumber\\
    & & \\
    K_B(x,y) &=& \dfrac{3+\xi}{4\pi}\left[
      \dfrac{\alpha_s(x)}{x}\theta(x-y) + (x\leftrightarrow y)
    \right].
  \nonumber\\
    & & 
  \end{eqnarray}
\end{subequations}

It is known that the existence of non-vanishing correction to the wave
function $A(x)$ makes the ladder approximation inconsistent with the
gauge symmetry.
We thus take Landau gauge $\xi=0$ here in our numerical calculation of
the improved ladder SD equation.

The quark mass function $B(x)$ can be regarded as
an order parameter of the chiral phase transition in QCD\@.
Non-zero solution of the SD equation $B(x)$ thus indicates dynamical
chiral symmetry breaking and implies the appearance of the
Nambu-Goldstone (NG) boson, the pion.
The decay constant of pion $f_\pi$ can be related with the mass
function $B(x)$.

In the following numerical analysis, we use the Pagels-Stokar formula
for the pion decay constant~\cite{Pagels:hd},
\begin{equation}
  f_{\rm PS}^2 = \dfrac{N_c}{4\pi^2}\int_0^{\Lambda^2} dx x
    \dfrac{B^2(x) - \dfrac{x}{4} \dfrac{d}{dx} B^2(x)}
          {\left( x + B^2(x) \right)^2}.
\label{eq:pagels-stokar}
\end{equation}

The Pagels-Stokar formula Eq.(\ref{eq:pagels-stokar}), however, does
not fully take account of the pion wave-function of the ladder
approximation.
The formula is therefore not ladder-exact one.
We thus relate it with the ladder-exact $f_\pi$ by
\begin{equation}
  f_\pi = N_{\rm PS} f_{\rm PS}.
\label{eq:nps}
\end{equation}
Fortunately, the Pagels-Stokar formula is known to work
extremely well within the Higashijima-Miransky approximation of the
ladder SD equation.~\cite{Aoki:1990eq}
We thus assume
\begin{equation}
  N_{\rm PS} \simeq 1,
\end{equation}
in this section.

\begin{figure*}[htbp]
  \begin{center}
    \begin{minipage}{0.45\textwidth}
      \begin{center}
        \psfrag{(a)}[][]{{\LARGE {\bf (a)}}}
        \psfrag{alp1/alp0}[][]{{\LARGE $\alpha_1/\alpha_0$}}
        \psfrag{fpi}[][]{{\Large $f_\pi$ [MeV]}}
        \psfrag{MS 1loop}[][]{{\large $\overline{\rm MS}$ one-loop}}
        \psfrag{MS 2loop}[][]{{\large $\overline{\rm MS}$ two-loop}}
        \psfrag{fps}[][]{{\LARGE $f_{\rm PS}$ [MeV]}}
        \psfrag{alp0/pi=2}[][]{{\large $\alpha_0/\pi=2$}}
        \psfrag{alp0/pi=3}[][]{{\large \phantom{$\alpha_0/\pi$}${}=3$}}
        \psfrag{alp0/pi=4}[][]{{\large \phantom{$\alpha_0/\pi$}${}=4$}}
        \psfrag{alp0/pi=5}[][]{{\large $\alpha_0/\pi=5$}}
        \psfrag{alp0/pi=6}[][]{{\large $\alpha_0/\pi=6$}}
        \psfrag{alp0/pi=7}[][]{{\large $\alpha_0/\pi=7$}}
        \psfrag{alp0/pi=8}[][]{{\large $\alpha_0/\pi=8$}}
        \includegraphics[width=6.5cm]{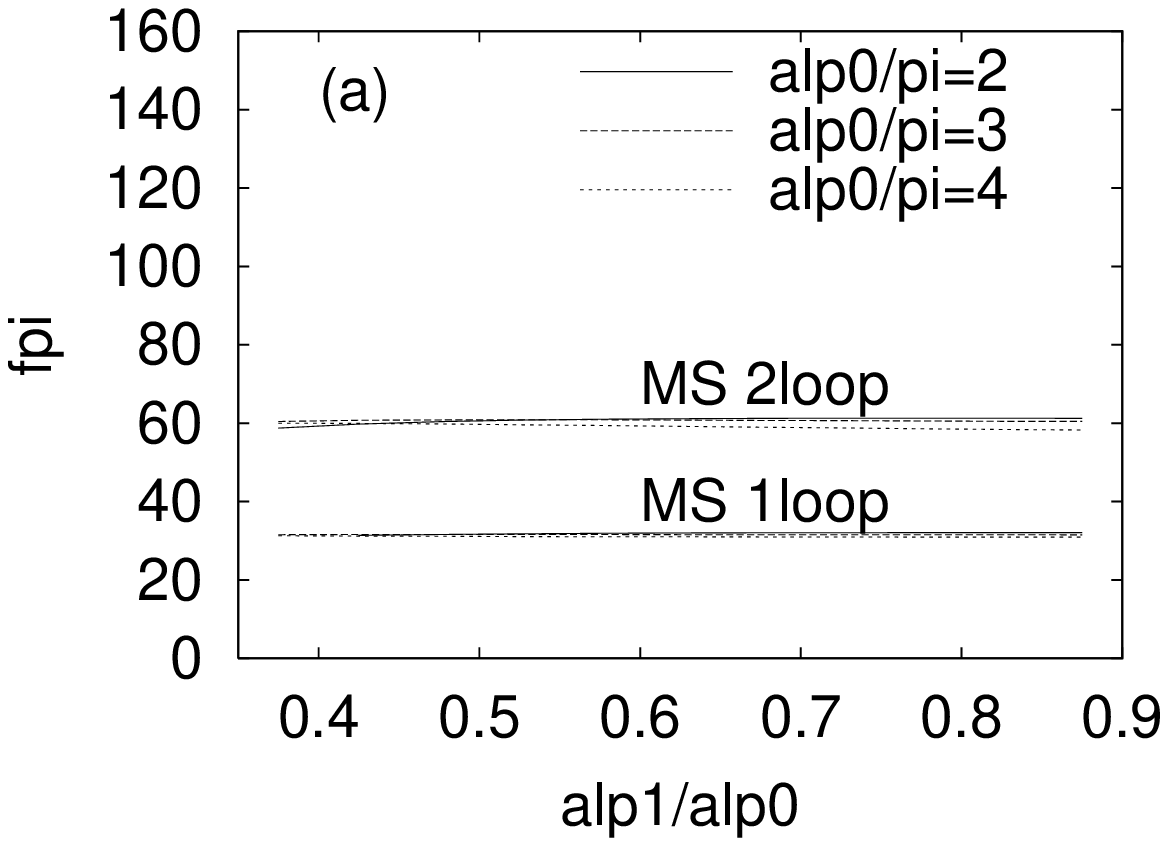}
      \end{center}
    \end{minipage}
    \begin{minipage}{0.45\textwidth}
      \begin{center}
        \psfrag{(b)}[][]{{\LARGE {\bf (b)}}}
        \psfrag{alp1/alp0}[][]{{\LARGE $\alpha_1/\alpha_0$}}
        \psfrag{mconst}[][]{{\Large $m_{\rm const}$ [MeV]}}
        \psfrag{MS 1loop}[][]{{\large $\overline{\rm MS}$ one-loop}}
        \psfrag{MS 2loop}[][]{{\large $\overline{\rm MS}$ two-loop}}
        \psfrag{alp0/pi=2}[][]{{\large $\alpha_0/\pi=2$}}
        \psfrag{alp0/pi=3}[][]{{\large \phantom{$\alpha_0/\pi$}${}=3$}}
        \psfrag{alp0/pi=4}[][]{{\large \phantom{$\alpha_0/\pi$}${}=4$}}
        \psfrag{alp0/pi=5}[][]{{\large $\alpha_0/\pi=5$}}
        \psfrag{alp0/pi=6}[][]{{\large $\alpha_0/\pi=6$}}
        \psfrag{alp0/pi=7}[][]{{\large $\alpha_0/\pi=7$}}
        \psfrag{alp0/pi=8}[][]{{\large $\alpha_0/\pi=8$}}
        \includegraphics[width=6.5cm]{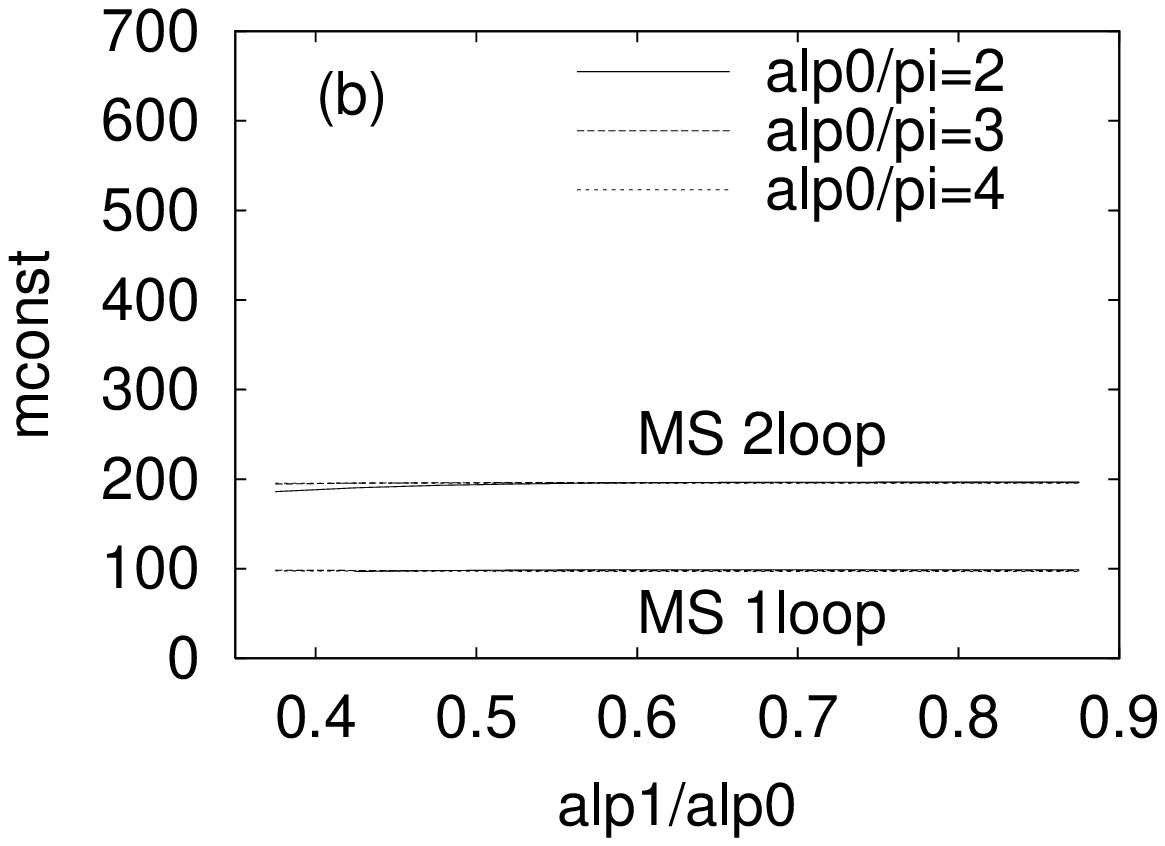}
      \end{center}
    \end{minipage}
  \end{center}
\caption{Results of {\bf (a)} the pion decay constant $f_\pi$  and 
  {\bf (b)} the ``constituent quark mass'' $m_{\rm const}$ with use of the 
  one- and two-loop $\overline{\rm MS}$ couplings.  The Higashijima-Miransky
  approximation is adopted in the kernel of the ladder SD equation.
  The boundary condition of RGE is assumed to be 
  $\alpha_s^{\overline{\rm MS}}(m_b)=0.2197$, which  corresponds to
  $\alpha_s^{\overline{\rm MS}}(M_Z)=0.1172$ at the scale of $M_Z$.
  $N_{\rm PS}$ is taken to be 1.
}
\label{fig:hig-msbar}
\end{figure*}

We are now ready to perform numerical analyses of the improved ladder
SD equation.
The UV cutoff $\Lambda$ is taken to be $\Lambda=m_b$.
We adopt an algorithm similar to the one described in
Ref.\cite{Hashimoto:2000uk} in our
numerical analysis.

Let us first confirm previous results of
Ref.\cite{Aoki:1990eq,Jain:1991pk}, where 
the momentum dependence of $\alpha_s(z)$ is given by the
$\overline{\rm MS}$ coupling with the renormalization scale $\mu$
being naively identified with $\sqrt{z}$.
Figure~\ref{fig:hig-msbar}a shows the result of $f_\pi$ for the one-
and two-loop $\overline{\rm MS}$ coupling.
We note that the calculated $f_\pi$ is extremely stable over wide
range of the IR coupling regularization parameters $\alpha_0$ and
$\alpha_1$. 
The result is also consistent with previous one that the calculated
$f_\pi$ is significantly smaller than its experimentally observed
value $f_\pi = 92.4$MeV\@.

As we expect from the running behavior of $\alpha_s$ 
(Figure~\ref{fig:running}), the $f_\pi$ calculated from two-loop 
$\overline{\rm MS}$ $\alpha_s$ is larger than the value from the
one-loop $\overline{\rm MS}$ coupling.
Although three- or higher-loop running effects may further enhance the
value of $f_\pi$, we expect that they are not enough to reproduce the
value $f_\pi=92.4$MeV, since the higher-loop $\overline{\rm MS}$
coupling is still well below the effective coupling at the scale 
$\sim 700$MeV\@, where the dynamical
chiral symmetry breaking takes place in the analysis of the ladder SD
equation.~\cite{Kugo:1992zg} 

We also emphasize that the mass function at the zero momentum $B(0)$
depends significantly on the choice of the IR parameters $\alpha_0$
and $\alpha_1$.
We find, however, that the ``constituent quark mass'' defined
as~\cite{Aoki:1990eq} 
\begin{equation}
  m_{\rm const} = B(4m_{\rm const}^2),
\end{equation}
is rather insensitive to $\alpha_0$ and $\alpha_1$. 
(See Figure~~\ref{fig:hig-msbar}b.)
Our result on $m_{\rm const}$ is also consistent with the previous
analysis.

\begin{figure*}[htbp]
  \begin{center}
    \begin{minipage}{0.45\textwidth}
      \begin{center}
        \psfrag{(a)}[][]{{\LARGE {\bf (a)}}}
        \psfrag{alp1/alp0}[][]{{\LARGE $\alpha_1/\alpha_0$}}
        \psfrag{fpi}[][]{{\Large $f_\pi$ [MeV]}}
        \psfrag{xi=0}[][]{{\large $\xi_{\rm bg}=0$}}
        \psfrag{xi=1}[][]{{\large $\xi_{\rm bg}=1$}}
        \psfrag{alp0/pi=2}[][]{{\large $\alpha_0/\pi=2$}}
        \psfrag{alp0/pi=3}[][]{{\large \phantom{$\alpha_0/\pi$}${}=3$}}
        \psfrag{alp0/pi=4}[][]{{\large \phantom{$\alpha_0/\pi$}${}=4$}}
        \psfrag{alp0/pi=5}[][]{{\large $\alpha_0/\pi=5$}}
        \psfrag{alp0/pi=6}[][]{{\large $\alpha_0/\pi=6$}}
        \psfrag{alp0/pi=7}[][]{{\large $\alpha_0/\pi=7$}}
        \psfrag{alp0/pi=8}[][]{{\large $\alpha_0/\pi=8$}}
        \includegraphics[width=6.5cm]{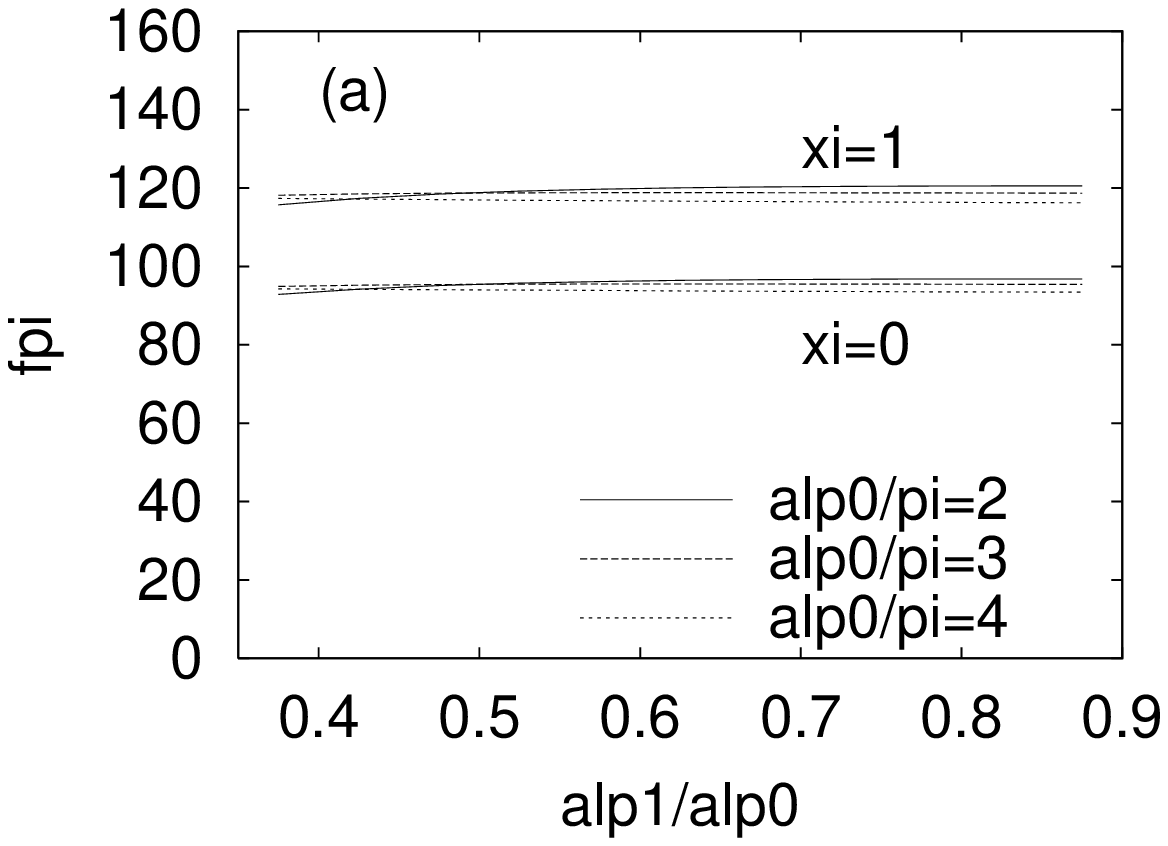}
      \end{center}
    \end{minipage}
    \begin{minipage}{0.45\textwidth}
      \begin{center}
        \psfrag{(b)}[][]{{\LARGE {\bf (b)}}}
        \psfrag{alp1/alp0}[][]{{\LARGE $\alpha_1/\alpha_0$}}
        \psfrag{mconst}[][]{{\Large $m_{\rm const}$ [MeV]}}
        \psfrag{xi=0}[][]{{\large $\xi_{\rm bg}=0$}}
        \psfrag{xi=1}[][]{{\large $\xi_{\rm bg}=1$}}
        \psfrag{alp0/pi=2}[][]{{\large $\alpha_0/\pi=2$}}
        \psfrag{alp0/pi=3}[][]{{\large \phantom{$\alpha_0/\pi$}${}=3$}}
        \psfrag{alp0/pi=4}[][]{{\large \phantom{$\alpha_0/\pi$}${}=4$}}
        \psfrag{alp0/pi=5}[][]{{\large $\alpha_0/\pi=5$}}
        \psfrag{alp0/pi=6}[][]{{\large $\alpha_0/\pi=6$}}
        \psfrag{alp0/pi=7}[][]{{\large $\alpha_0/\pi=7$}}
        \psfrag{alp0/pi=8}[][]{{\large $\alpha_0/\pi=8$}}
        \includegraphics[width=6.5cm]{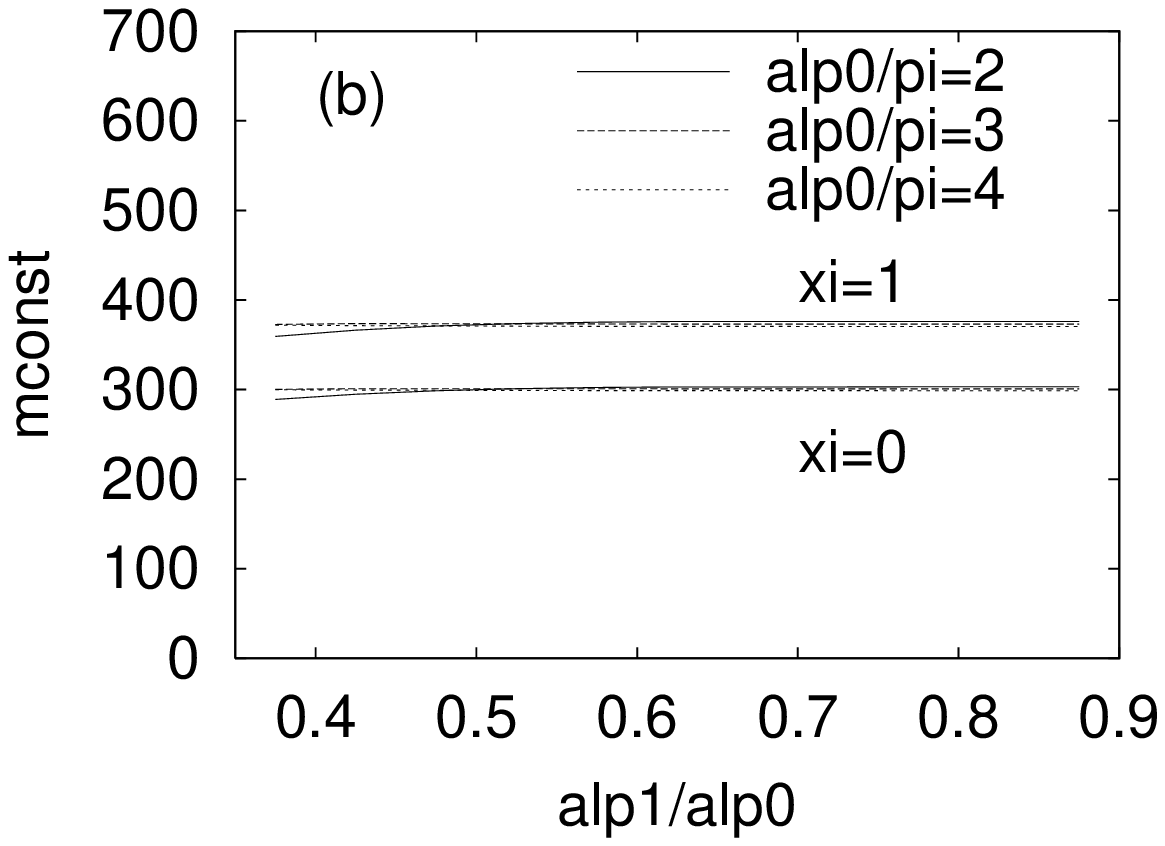}
      \end{center}
    \end{minipage}
  \end{center}
\caption{Results of {\bf (a)} the pion decay constant $f_\pi$  and 
  {\bf (b)} the ``constituent quark mass'' $m_{\rm const}$ with use of the 
  effective couplings (two-loop RGE + finite part).
  The cases with $\xi_{\rm bg}=0$ and $\xi_{\rm bg}=1$ are shown.
  The Higashijima-Miransky
  approximation is adopted in the kernel of the ladder SD equation.
  The boundary condition of RGE is assumed to be 
  $\alpha_s^{\overline{\rm MS}}(m_b)=0.2197$, which  corresponds to
  $\alpha_s^{\overline{\rm MS}}(M_Z)=0.1172$ at the scale of $M_Z$.
  $N_{\rm PS}$ is taken to be 1.
}
\label{fig:hig-eff2}
\end{figure*}

We next consider the improved ladder SD equation combined with the
effective coupling Eq.(\ref{eq:geffhigh}) and Eq.(\ref{eq:geff}).
We adopt two-loop $\overline{\rm MS}$ coupling as
$\alpha_s^{\overline{\rm MS}}(\mu)$ in Eq.(\ref{eq:geffhigh}) and
Eq.(\ref{eq:geff}). 
The leading-log dependence on $\mu$ cancels in the definition of
$\alpha_s^{\rm eff}$ as we noted before.
We take $\alpha_s^{\rm eff}(-k^2=z, \mu=\sqrt{z})$ in order to take
account of the next-to-leading-log effects.

We show our results on $f_\pi$ and $m_{\rm const}$ in
Figure~\ref{fig:hig-eff2}. 
Again, the results are quite insensitive to the variation of IR
regularization parameters $\alpha_0$ and $\alpha_1$.
We also note that the calculated $f_\pi$ from $\alpha_s^{\rm eff}$
with $\xi_{\rm bg}=0$ agrees very well with its experimentally
observed value $f_\pi=92.4$MeV\@.
The result suggests the validity of our method to settle the scale
ambiguity in the improved ladder SD equation.
It may also imply importance of the ladder diagrams in the mechanism
of the dynamical chiral symmetry breaking of QCD\@.

If we adopt $\xi_{\rm bg}=1$ in the effective coupling, on the other
hand, the calculated $f_\pi$ becomes sizably larger than 92.4MeV (of
order 120MeV). 
This deviation of $f_\pi$ in $\xi_{\rm bg}=1$ may be understood as an
indication of the double counting of the ladder diagrams as we
described before.
This point should be investigated further in future.

\begin{figure*}[htbp]
  \begin{center}
    \begin{minipage}{0.45\textwidth}
      \begin{center}
        \psfrag{(a)}[][]{{\LARGE {\bf (a)}}}
        \psfrag{alp1/alp0}[][]{{\LARGE $\alpha_1/\alpha_0$}}
        \psfrag{fpi}[][]{{\Large $f_\pi$ [MeV]}}
        \psfrag{xi=0}[][]{{\large $\xi_{\rm bg}=0$}}
        \psfrag{xi=1}[][]{{\large $\xi_{\rm bg}=1$}}
        \psfrag{alp0/pi=2}[][]{{\large $\alpha_0/\pi=2$}}
        \psfrag{alp0/pi=3}[][]{{\large \phantom{$\alpha_0/\pi$}${}=3$}}
        \psfrag{alp0/pi=4}[][]{{\large \phantom{$\alpha_0/\pi$}${}=4$}}
        \psfrag{alp0/pi=5}[][]{{\large $\alpha_0/\pi=5$}}
        \psfrag{alp0/pi=6}[][]{{\large $\alpha_0/\pi=6$}}
        \psfrag{alp0/pi=7}[][]{{\large $\alpha_0/\pi=7$}}
        \psfrag{alp0/pi=8}[][]{{\large $\alpha_0/\pi=8$}}
        \includegraphics[width=6.5cm]{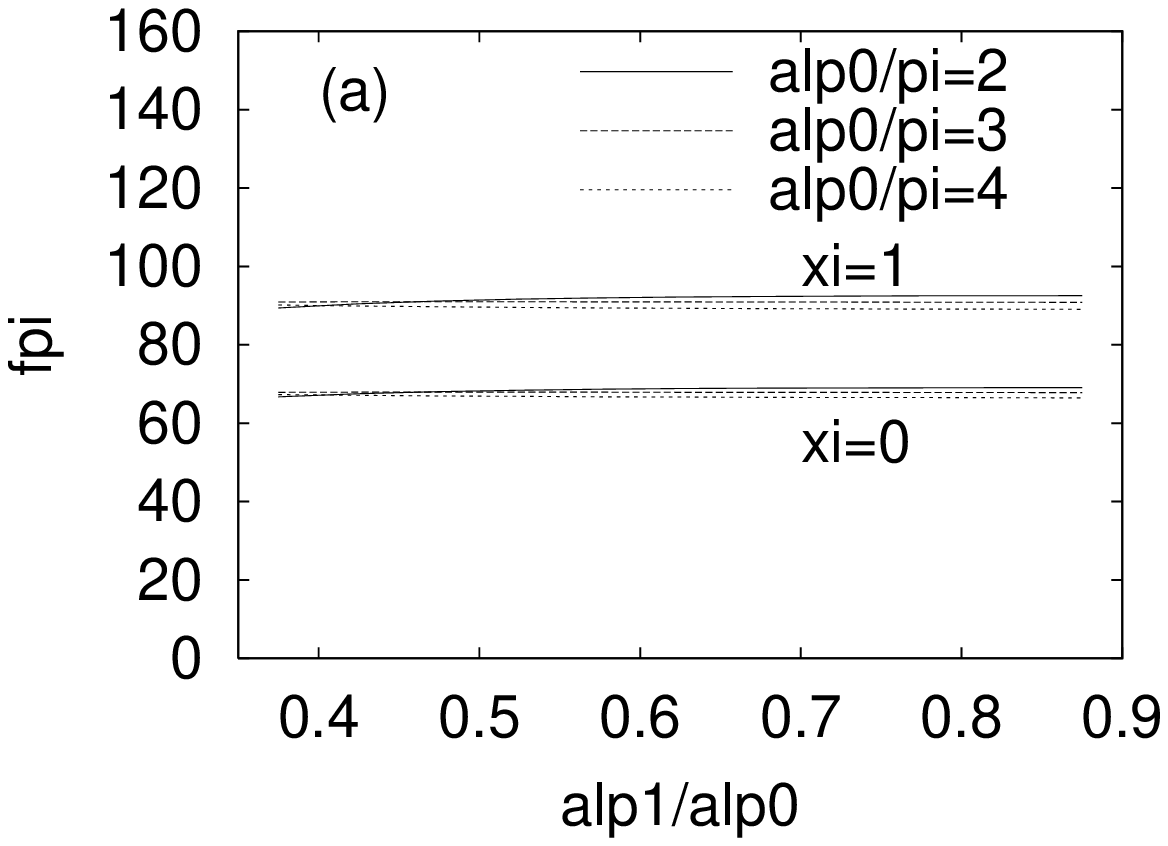}
      \end{center}
    \end{minipage}
    \begin{minipage}{0.45\textwidth}
      \begin{center}
        \psfrag{(b)}[][]{{\LARGE {\bf (b)}}}
        \psfrag{alp1/alp0}[][]{{\LARGE $\alpha_1/\alpha_0$}}
        \psfrag{mconst}[][]{{\Large $m_{\rm const}$ [MeV]}}
        \psfrag{xi=0}[][]{{\large $\xi_{\rm bg}=0$}}
        \psfrag{xi=1}[][]{{\large $\xi_{\rm bg}=1$}}
        \psfrag{alp0/pi=2}[][]{{\large $\alpha_0/\pi=2$}}
        \psfrag{alp0/pi=3}[][]{{\large \phantom{$\alpha_0/\pi$}${}=3$}}
        \psfrag{alp0/pi=4}[][]{{\large \phantom{$\alpha_0/\pi$}${}=4$}}
        \psfrag{alp0/pi=5}[][]{{\large $\alpha_0/\pi=5$}}
        \psfrag{alp0/pi=6}[][]{{\large $\alpha_0/\pi=6$}}
        \psfrag{alp0/pi=7}[][]{{\large $\alpha_0/\pi=7$}}
        \psfrag{alp0/pi=8}[][]{{\large $\alpha_0/\pi=8$}}
        \includegraphics[width=6.5cm]{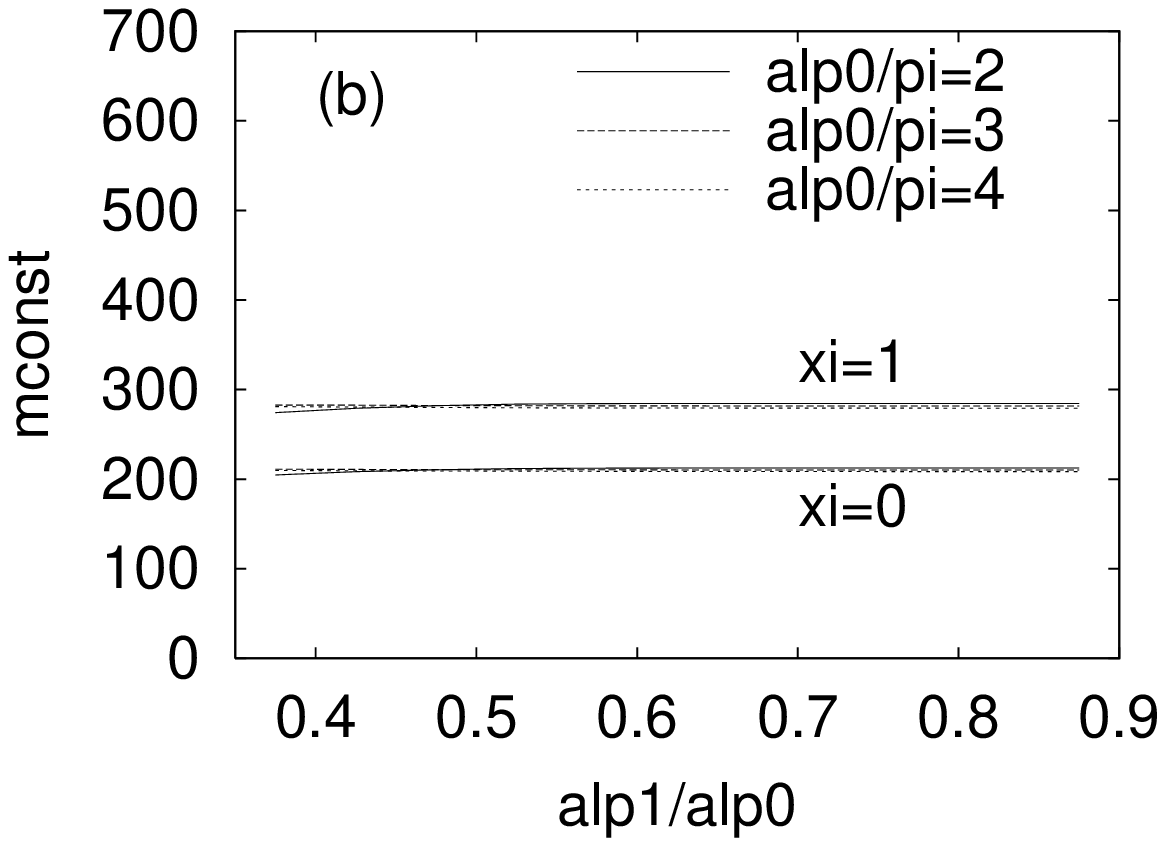}
      \end{center}
    \end{minipage}
  \end{center}
\caption{Results of {\bf (a)} the pion decay constant $f_\pi$  and 
  {\bf (b)} the ``constituent quark mass'' $m_{\rm const}$ with use of the 
  effective couplings neglecting their two-loop RGE effects.
  The cases with $\xi_{\rm bg}=0$ and $\xi_{\rm bg}=1$ are shown.
  The Higashijima-Miransky
  approximation is adopted in the kernel of the ladder SD equation.
  The boundary condition of RGE is assumed to be 
  $\alpha_s^{\overline{\rm MS}}(m_b)=0.2197$, which  corresponds to
  $\alpha_s^{\overline{\rm MS}}(M_Z)=0.1172$ at the scale of $M_Z$.
  $N_{\rm PS}$ is taken to be 1.
}
\label{fig:hig-eff1}
\end{figure*}

\begin{figure*}[htbp]
  \begin{center}
    \begin{minipage}{0.45\textwidth}
      \begin{center}
        \includegraphics[width=6cm]{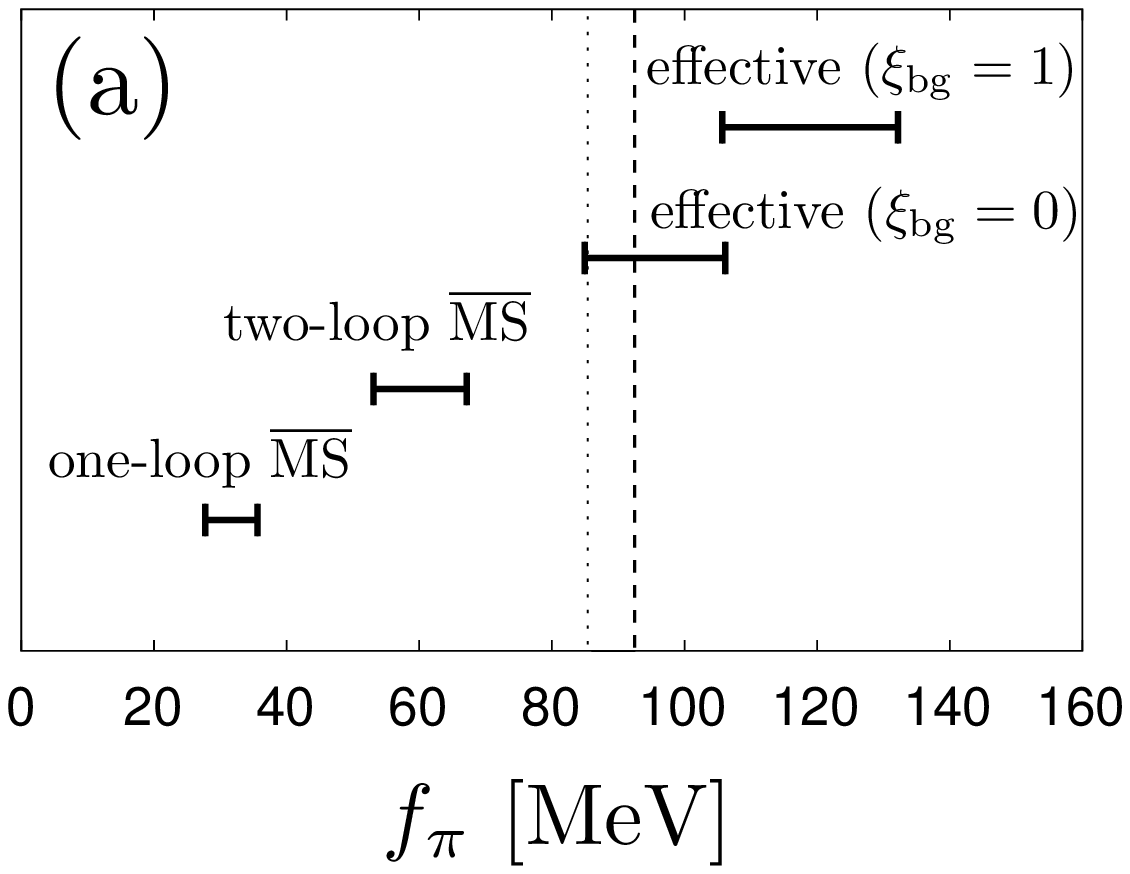}
      \end{center}
    \end{minipage}
    \begin{minipage}{0.45\textwidth}
      \begin{center}
        \includegraphics[width=6cm]{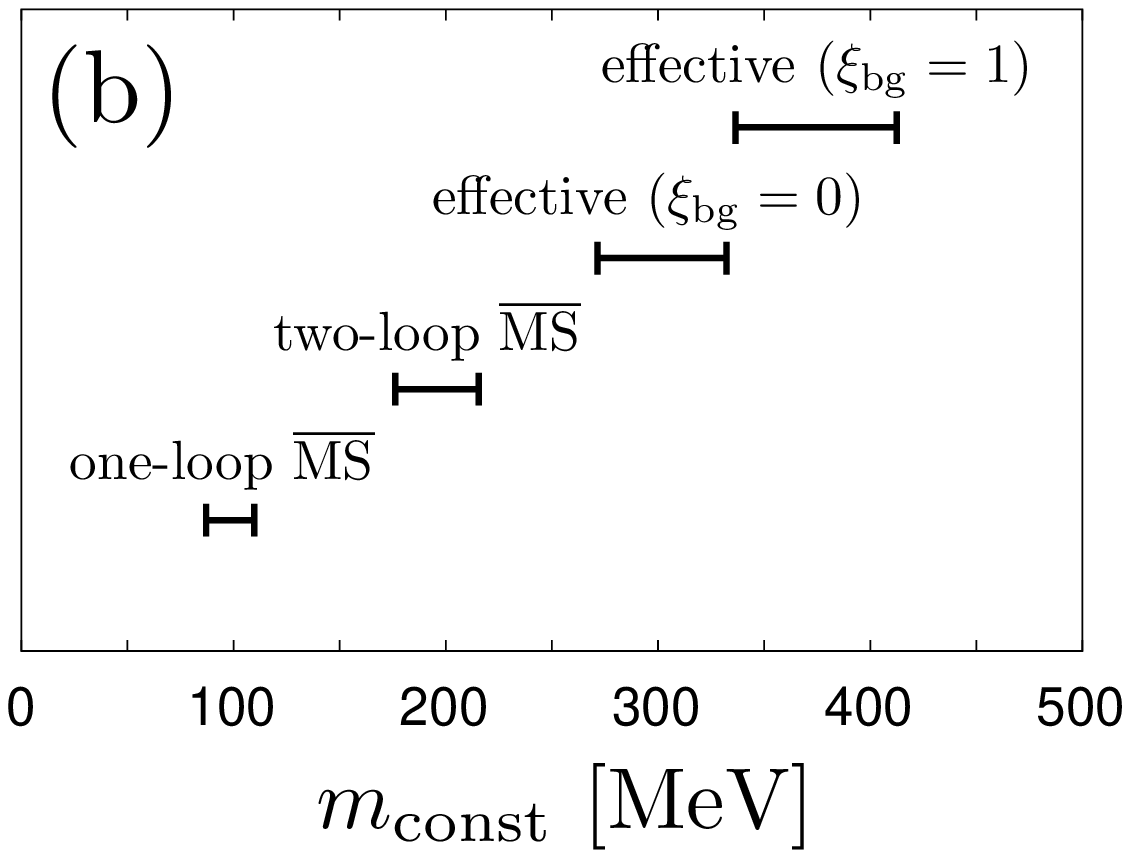}
      \end{center}
    \end{minipage}
  \end{center}
\caption{Uncertainties of our results on 
  {\bf (a)} the pion decay constant $f_\pi$ and 
  {\bf (b)} the ``constituent quark mass'' $m_{\rm const}$.
  The strong coupling constant $\alpha_s$ is taken in range of
  $\alpha_s^{\overline{\rm MS}}(m_b)=0.2197\pm 0.0075$, which
  corresponds to $\alpha_s^{\overline{\rm MS}}(M_Z)=0.1172\pm 0.0020$
  at the scale of $M_Z$. 
  The IR regularization $\alpha_0$ and $\alpha_1$ are
  $2\pi<\alpha_0<4\pi$ and $0.4<\alpha_1/\alpha_0<0.8$.
  $N_{\rm PS}$ is taken to be 1.
  The vertical lines in {\bf (a)} indicate $f_\pi=92.4$MeV (the
  experimental value of $f_\pi$) and $f_\pi=86$MeV (the decay constant
  in the chiral limit). 
}
\label{fig:hig-maxmin}
\end{figure*}

In order to figure out the effects of the two-loop RGE in the
effective coupling, we next perform analyses using one-loop 
$\alpha_s^{\overline{\rm MS}}(\mu)$ in Eq.(\ref{eq:geffhigh}) and
Eq.(\ref{eq:geff}). 
In this case, the $\mu$-dependence is completely canceled in the
effective coupling. 
Our results are shown in Figure~\ref{fig:hig-eff1}.
These results are sizably smaller than the results of
Figure~\ref{fig:hig-eff2}.
The effect of two-loop RGE is therefore sizable in the effective
coupling. 
Although the $f_\pi$ at $\xi_{\rm bg}=1$ agrees well
with its experimental value 92.4MeV\@, 
the agreement should be regarded as an accidental coincidence.
Actually, the results shown in Figure~\ref{fig:hig-eff1} sizably
decrease if we adopt Eq.(\ref{eq:alphas_mz}) directly, instead of 
Eq.(\ref{eq:bc_rge}), as the boundary condition of the one-loop RGE\@.  

We have so far used the central value of Eq.(\ref{eq:bc_rge}), 
$\alpha_s^{\overline{\rm MS}}(m_b)=0.2197$ as an input parameter of
the QCD coupling.
Larger the $\alpha_s^{\overline{\rm MS}}(m_b)$, larger $f_\pi$
is obtained in our framework, however.
The uncertainty of the high energy QCD coupling thus leads to an 
uncertainty of our results on $f_\pi$.
There also exists uncertainty coming from the choice of the IR
regularization parameters $\alpha_0$ and $\alpha_1$.
Since our results (Figures~\ref{fig:hig-msbar}--\ref{fig:hig-eff1})
are rather insensitive to $\alpha_0$ and $\alpha_1$, however, 
the bulk of uncertainty comes from the high energy QCD coupling here
within the Higashijima-Miransky approximation. 

We show the uncertainty of our results in Figure~\ref{fig:hig-maxmin}.
The high energy QCD coupling is taken in range of
Eq.(\ref{eq:bc_rge}). 
The IR regularization parameters are also varied in range of 
$2\pi < \alpha_0 < 4\pi$, $0.4 < \alpha_1/\alpha_0 < 0.8$ in the
plot.
The experimentally observed value of $f_\pi=92.4$MeV and the decay
constant in the chiral limit $f_\pi=86$MeV are also plotted in
Figure~\ref{fig:hig-maxmin}.
It is very impressing that the calculated value of $f_\pi$ agrees well
with its experimental value for the effective coupling with 
$\xi_{\rm bg}=0$. 

\section{Calculating $f_\pi$ with Non-local gauge}
\label{sec:non-local}

We next try to perform the angular integral of the kernel 
Eq.(\ref{eq:kernel}) numerically without using the approximation 
Eq.(\ref{eq:higashijima-type}).
Unfortunately, the function $A(x)$ receives non-vanishing correction
even if we take $\xi=0$ in this case.
In order to keep $A\equiv 1$, we need to use so called ``non-local
gauge'' fixing in which the gauge parameter $\xi$ is regarded as a
function of the gluon
momentum.~\cite{Georgi:1989cd,Kugo:1992pr}

It is known that $A\equiv 1$ can be achieved when we take a particular
form~\cite{Kugo:1992pr} of function $\xi$,
\begin{equation}
  \xi(z) = \dfrac{3}{z^2 \alpha_s(z)} 
  \int_0^z dz z^2 \frac{d}{dz} \alpha_s(z).
\label{eq:non-local}  
\end{equation}
We then obtain
\begin{subequations}
  \begin{eqnarray}
    K_A(x,y) &=& 0, \\
    K_B(x,y) &=& \dfrac{1}{2\pi^2}\int_0^\pi \!\!\! d\theta 
       \sin^2\theta \dfrac{(3+\xi(z))\alpha_s(z)}{z},
    \nonumber\\
    & &
  \end{eqnarray}
\end{subequations}
with $z\equiv x+y-2\sqrt{xy}\cos\theta$.

\begin{figure*}[htbp]
  \begin{center}
    \begin{minipage}{0.45\textwidth}
      \begin{center}
        \psfrag{(a)}[][]{{\LARGE {\bf (a)}}}
        \psfrag{alp1/alp0}[][]{{\LARGE $\alpha_1/\alpha_0$}}
        \psfrag{fpi}[][]{{\Large $f_\pi$ [MeV]}}
        \psfrag{MS 1loop}[][]{{\large $\overline{\rm MS}$ one-loop}}
        \psfrag{MS 2loop}[][]{{\large $\overline{\rm MS}$ two-loop}}
        \psfrag{fps}[][]{{\LARGE $f_{\rm PS}$ [MeV]}}
        \psfrag{alp0/pi=2}[][]{{\large $\alpha_0/\pi=2$}}
        \psfrag{alp0/pi=3}[][]{{\large \phantom{$\alpha_0/\pi$}${}=3$}}
        \psfrag{alp0/pi=4}[][]{{\large \phantom{$\alpha_0/\pi$}${}=4$}}
        \psfrag{alp0/pi=5}[][]{{\large $\alpha_0/\pi=5$}}
        \psfrag{alp0/pi=6}[][]{{\large $\alpha_0/\pi=6$}}
        \psfrag{alp0/pi=7}[][]{{\large $\alpha_0/\pi=7$}}
        \psfrag{alp0/pi=8}[][]{{\large $\alpha_0/\pi=8$}}
        \includegraphics[width=6.5cm]{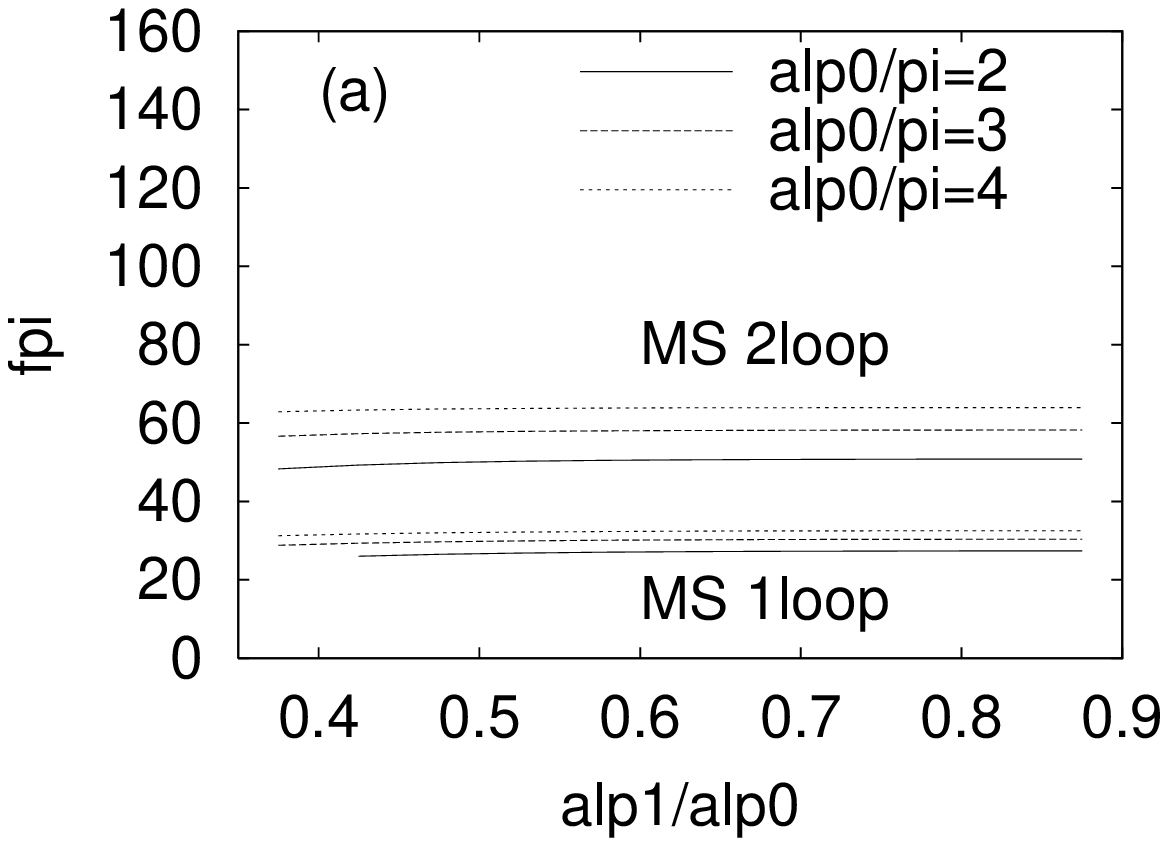}
      \end{center}
    \end{minipage}
    \begin{minipage}{0.45\textwidth}
      \begin{center}
        \psfrag{(b)}[][]{{\LARGE {\bf (b)}}}
        \psfrag{alp1/alp0}[][]{{\LARGE $\alpha_1/\alpha_0$}}
        \psfrag{mconst}[][]{{\Large $m_{\rm const}$ [MeV]}}
        \psfrag{MS 1loop}[][]{{\large $\overline{\rm MS}$ one-loop}}
        \psfrag{MS 2loop}[][]{{\large $\overline{\rm MS}$ two-loop}}
        \psfrag{alp0/pi=2}[][]{{\large $\alpha_0/\pi=2$}}
        \psfrag{alp0/pi=3}[][]{{\large \phantom{$\alpha_0/\pi$}${}=3$}}
        \psfrag{alp0/pi=4}[][]{{\large \phantom{$\alpha_0/\pi$}${}=4$}}
        \psfrag{alp0/pi=5}[][]{{\large $\alpha_0/\pi=5$}}
        \psfrag{alp0/pi=6}[][]{{\large $\alpha_0/\pi=6$}}
        \psfrag{alp0/pi=7}[][]{{\large $\alpha_0/\pi=7$}}
        \psfrag{alp0/pi=8}[][]{{\large $\alpha_0/\pi=8$}}
        \includegraphics[width=6.5cm]{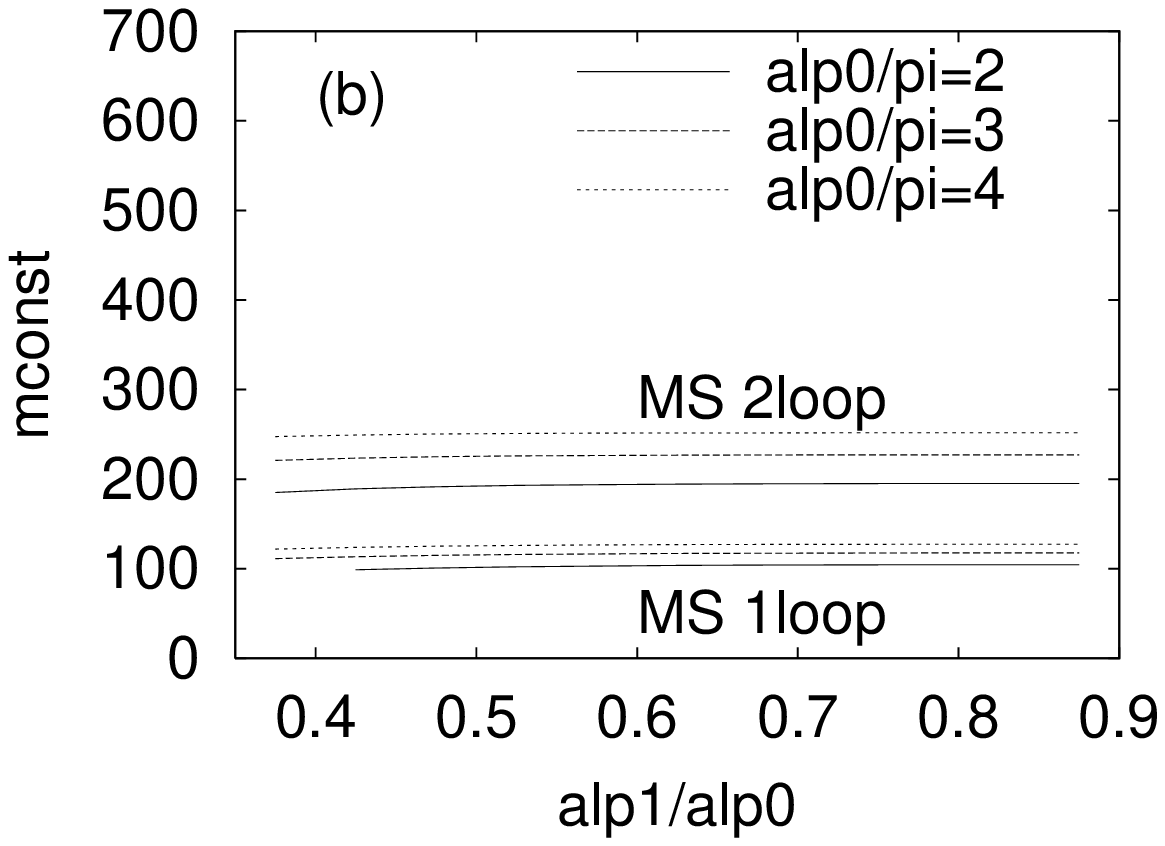}
      \end{center}
    \end{minipage}
  \end{center}
\caption{Results of {\bf (a)} the pion decay constant $f_\pi$  and 
  {\bf (b)} the ``constituent quark mass'' $m_{\rm const}$ with use of the 
  one- and two-loop $\overline{\rm MS}$ couplings.  The non-local
  gauge fixing method is adopted in the ladder SD equation.
  The boundary condition of RGE is assumed to be 
  $\alpha_s^{\overline{\rm MS}}(m_b)=0.2197$, which  corresponds to
  $\alpha_s^{\overline{\rm MS}}(M_Z)=0.1172$ at the scale of $M_Z$.
  $N_{\rm PS}$ is taken to be $0.8$.
}
\label{fig:gap-msbar}
\end{figure*}

We first start with naive improved ladder case, in which
$\alpha_s^{\overline{\rm MS}}(\mu=\sqrt{z})$ is used as the running
coupling of the SD equation.
The naive improved ladder SD equation with one-loop
$\alpha_s^{\overline{\rm MS}}$ has been investigated in full detail by
Ref.\cite{Kugo:1992zg}.
Unlike the case with Higashijima-Miransky approximation, it was found that the
Pagels-Stokar formula Eq.(\ref{eq:pagels-stokar}) gives sizably larger
$f_\pi$ than the ladder-exact value of $f_\pi$ in the non-local gauge
fixing method. 
The factor $N_{\rm PS}$ in Eq.(\ref{eq:nps}) is then smaller than
unity. 
We extract here the value of $N_{\rm PS}$ from Ref.\cite{Kugo:1992zg}
for various IR coupling parameters $\alpha_0$, $\alpha_1$.
The results are summarized in table~\ref{tab:nps}. 
We find $N_{\rm PS}$ is of order $0.8$ and it is a decreasing function
of $\alpha_0$, $\alpha_1$.
We assume here $N_{\rm PS}=0.8$ irrespective to $\alpha_0$,$\alpha_1$
throughout in this section.

\begin{table}[htbp]
  \centering
  \begin{tabular}{|rr|c|}
    \hline
    $\alpha_0/\pi$ & $\alpha_1/\pi$ & $N_{\rm PS}$ \\
    \hline
    $2.3$  & $0.74$ & $0.85$ \\
    $3.1$  & $0.89$ & $0.83$ \\
    $4.4$  & $1.1$ & $0.80$ \\
    $7.2$  & $1.5$ & $0.76$ \\
    \hline
  \end{tabular}
  \caption{The value of $N_{\rm PS}$ for various $\alpha_0$ and
    $\alpha_1$ with one-loop $\overline{\rm MS}$ RGE.} 
  \label{tab:nps}
\end{table}

Figure~\ref{fig:gap-msbar} shows our results of the naive improved
ladder SD equation with one- and two-loop $\overline{\rm MS}$
couplings. 
Our result of the one-loop $\overline{\rm MS}$ coupling is consistent
with the analysis of Ref.\cite{Kugo:1992zg}.
Although both $f_\pi$ and $m_{\rm const}$ are relatively stable over
the variation of $\alpha_1/\alpha_0$, their dependence on $\alpha_0$
is of non-negligible order.
We are thus not able to calculate $f_\pi$ in a reliable manner within
the non-local gauge fixing method.
The size of the calculated $f_\pi$ is, however, significantly smaller
than its experimental value 92.4MeV, even if we use two-loop
$\alpha_s^{\overline{\rm MS}}$ and assume very strong
coupling in the IR region $\alpha_0=4\pi$.
This fact again suggests the importance of the scale ambiguity in the
improved ladder SD equation.

It should be emphasized that 
$N_{\rm PS}=0.8$ is assumed irrespective to the value of $\alpha_0$ 
in Figure~\ref{fig:gap-msbar}.
Since the factor $N_{\rm PS}$ is a decreasing function of $\alpha_0$,
the ladder-exact value of $f_\pi$ is considered more stable than the
result of the Pagels-Stokar formula.
In order to make the analysis more reliable, we therefore need to
evaluate the ladder-exact value of $f_\pi$ using the non-local gauge
fixing method. 
Such a calculation is technically difficult to be performed, however.
We thus leave the subject as a problem to be examined in future.

\begin{figure*}[htbp]
  \begin{center}
    \begin{minipage}{0.45\textwidth}
      \begin{center}
        \psfrag{(a)}[][]{{\LARGE {\bf (a)}}}
        \psfrag{alp1/alp0}[][]{{\LARGE $\alpha_1/\alpha_0$}}
        \psfrag{fpi}[][]{{\Large $f_\pi$ [MeV]}}
        \psfrag{xi=0}[][]{{\large $\xi_{\rm bg}=0$}}
        \psfrag{xi=1}[][]{{\large $\xi_{\rm bg}=1$}}
        \psfrag{alp0/pi=2}[][]{{\large $\alpha_0/\pi=2$}}
        \psfrag{alp0/pi=3}[][]{{\large \phantom{$\alpha_0/\pi$}${}=3$}}
        \psfrag{alp0/pi=4}[][]{{\large \phantom{$\alpha_0/\pi$}${}=4$}}
        \psfrag{alp0/pi=5}[][]{{\large $\alpha_0/\pi=5$}}
        \psfrag{alp0/pi=6}[][]{{\large $\alpha_0/\pi=6$}}
        \psfrag{alp0/pi=7}[][]{{\large $\alpha_0/\pi=7$}}
        \psfrag{alp0/pi=8}[][]{{\large $\alpha_0/\pi=8$}}
        \includegraphics[width=6.5cm]{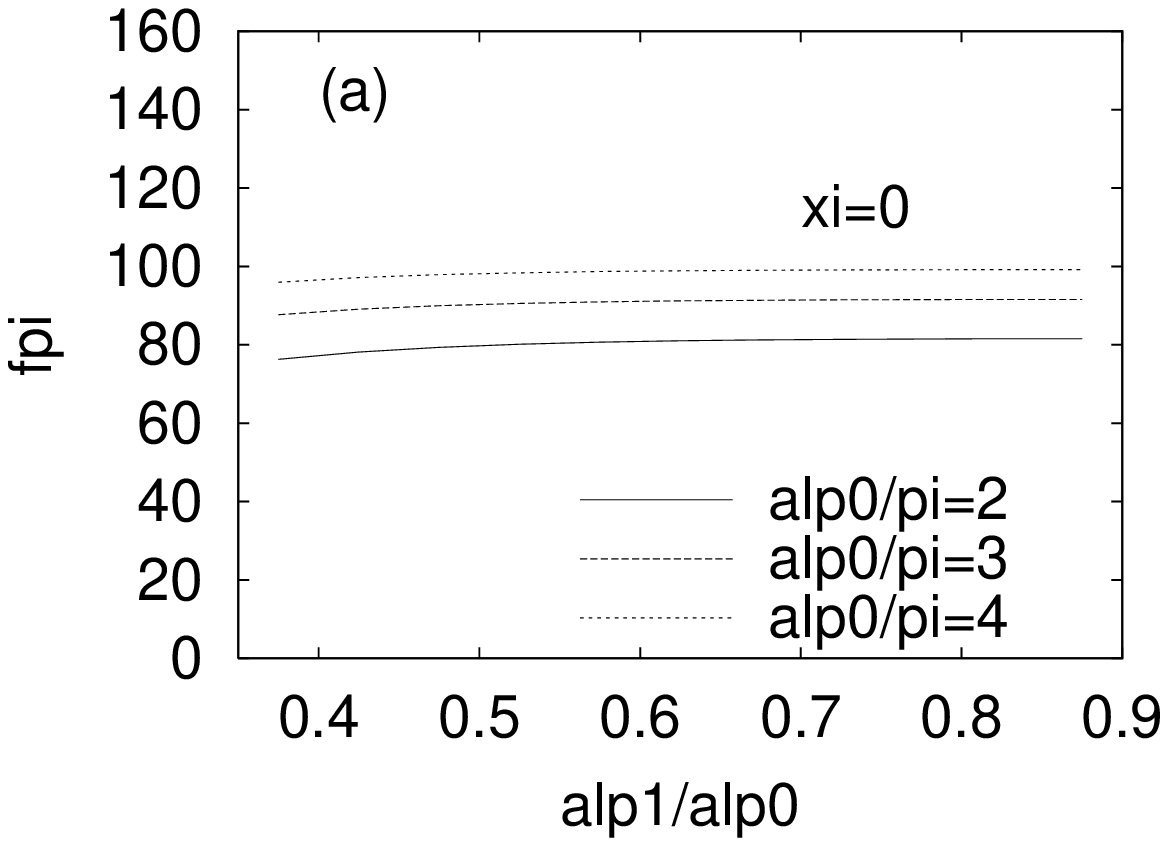}
      \end{center}
    \end{minipage}
    \begin{minipage}{0.45\textwidth}
      \begin{center}
        \psfrag{(b)}[][]{{\LARGE {\bf (b)}}}
        \psfrag{alp1/alp0}[][]{{\LARGE $\alpha_1/\alpha_0$}}
        \psfrag{mconst}[][]{{\Large $m_{\rm const}$ [MeV]}}
        \psfrag{xi=0}[][]{{\large $\xi_{\rm bg}=0$}}
        \psfrag{xi=1}[][]{{\large $\xi_{\rm bg}=1$}}
        \psfrag{alp0/pi=2}[][]{{\large $\alpha_0/\pi=2$}}
        \psfrag{alp0/pi=3}[][]{{\large \phantom{$\alpha_0/\pi$}${}=3$}}
        \psfrag{alp0/pi=4}[][]{{\large \phantom{$\alpha_0/\pi$}${}=4$}}
        \psfrag{alp0/pi=5}[][]{{\large $\alpha_0/\pi=5$}}
        \psfrag{alp0/pi=6}[][]{{\large $\alpha_0/\pi=6$}}
        \psfrag{alp0/pi=7}[][]{{\large $\alpha_0/\pi=7$}}
        \psfrag{alp0/pi=8}[][]{{\large $\alpha_0/\pi=8$}}
        \includegraphics[width=6.5cm]{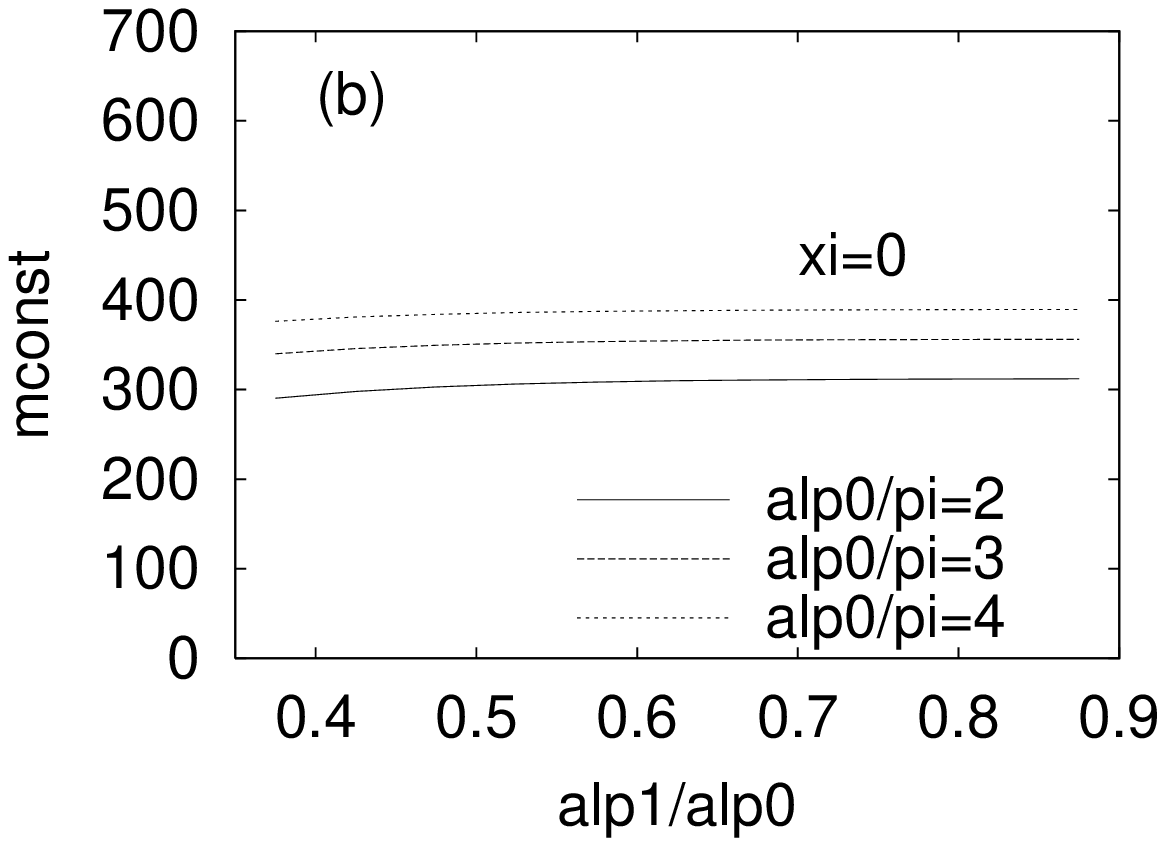}
      \end{center}
    \end{minipage}
  \end{center}
\caption{Results of {\bf (a)} the pion decay constant $f_\pi$  and 
  {\bf (b)} the ``constituent quark mass'' $m_{\rm const}$ with use of the 
  effective coupling (two-loop RGE + finite part) with 
  $\xi_{\rm bg}=0$.
  The non-local gauge fixing method is adopted in the ladder SD equation.
  The boundary condition of RGE is assumed to be 
  $\alpha_s^{\overline{\rm MS}}(m_b)=0.2197$, which  corresponds to
  $\alpha_s^{\overline{\rm MS}}(M_Z)=0.1172$ at the scale of $M_Z$.
  $N_{\rm PS}$ is taken to be $0.8$.
}
\label{fig:gap-eff0}
\end{figure*}

\begin{figure*}[htbp]
  \begin{center}
    \begin{minipage}{0.45\textwidth}
      \begin{center}
        \psfrag{(a)}[][]{{\LARGE {\bf (a)}}}
        \psfrag{alp1/alp0}[][]{{\LARGE $\alpha_1/\alpha_0$}}
        \psfrag{fpi}[][]{{\Large $f_\pi$ [MeV]}}
        \psfrag{xi=0}[][]{{\large $\xi_{\rm bg}=0$}}
        \psfrag{xi=1}[][]{{\large $\xi_{\rm bg}=1$}}
        \psfrag{alp0/pi=2}[][]{{\large $\alpha_0/\pi=2$}}
        \psfrag{alp0/pi=3}[][]{{\large \phantom{$\alpha_0/\pi$}${}=3$}}
        \psfrag{alp0/pi=4}[][]{{\large \phantom{$\alpha_0/\pi$}${}=4$}}
        \psfrag{alp0/pi=5}[][]{{\large $\alpha_0/\pi=5$}}
        \psfrag{alp0/pi=6}[][]{{\large $\alpha_0/\pi=6$}}
        \psfrag{alp0/pi=7}[][]{{\large $\alpha_0/\pi=7$}}
        \psfrag{alp0/pi=8}[][]{{\large $\alpha_0/\pi=8$}}
        \includegraphics[width=6.5cm]{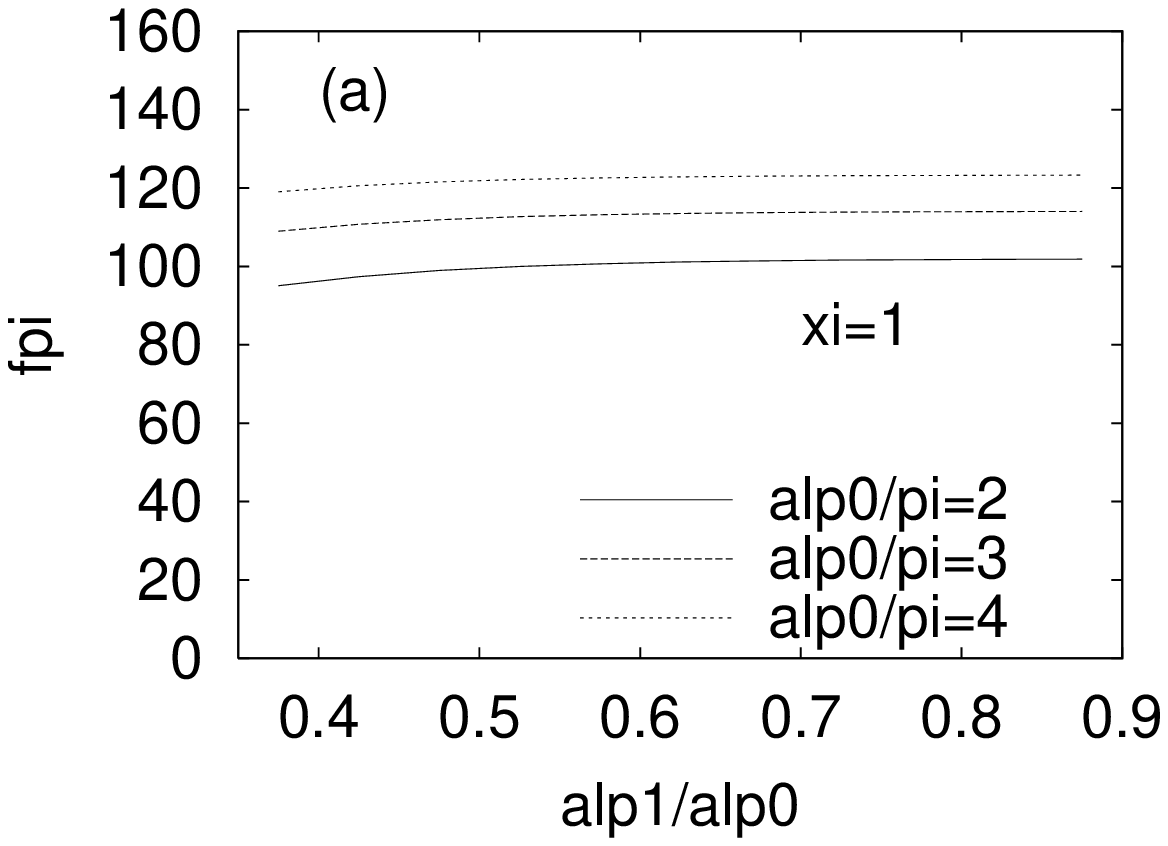}
      \end{center}
    \end{minipage}
    \begin{minipage}{0.45\textwidth}
      \begin{center}
        \psfrag{(b)}[][]{{\LARGE {\bf (b)}}}
        \psfrag{alp1/alp0}[][]{{\LARGE $\alpha_1/\alpha_0$}}
        \psfrag{mconst}[][]{{\Large $m_{\rm const}$ [MeV]}}
        \psfrag{xi=0}[][]{{\large $\xi_{\rm bg}=0$}}
        \psfrag{xi=1}[][]{{\large $\xi_{\rm bg}=1$}}
        \psfrag{alp0/pi=2}[][]{{\large $\alpha_0/\pi=2$}}
        \psfrag{alp0/pi=3}[][]{{\large \phantom{$\alpha_0/\pi$}${}=3$}}
        \psfrag{alp0/pi=4}[][]{{\large \phantom{$\alpha_0/\pi$}${}=4$}}
        \psfrag{alp0/pi=5}[][]{{\large $\alpha_0/\pi=5$}}
        \psfrag{alp0/pi=6}[][]{{\large $\alpha_0/\pi=6$}}
        \psfrag{alp0/pi=7}[][]{{\large $\alpha_0/\pi=7$}}
        \psfrag{alp0/pi=8}[][]{{\large $\alpha_0/\pi=8$}}
        \includegraphics[width=6.5cm]{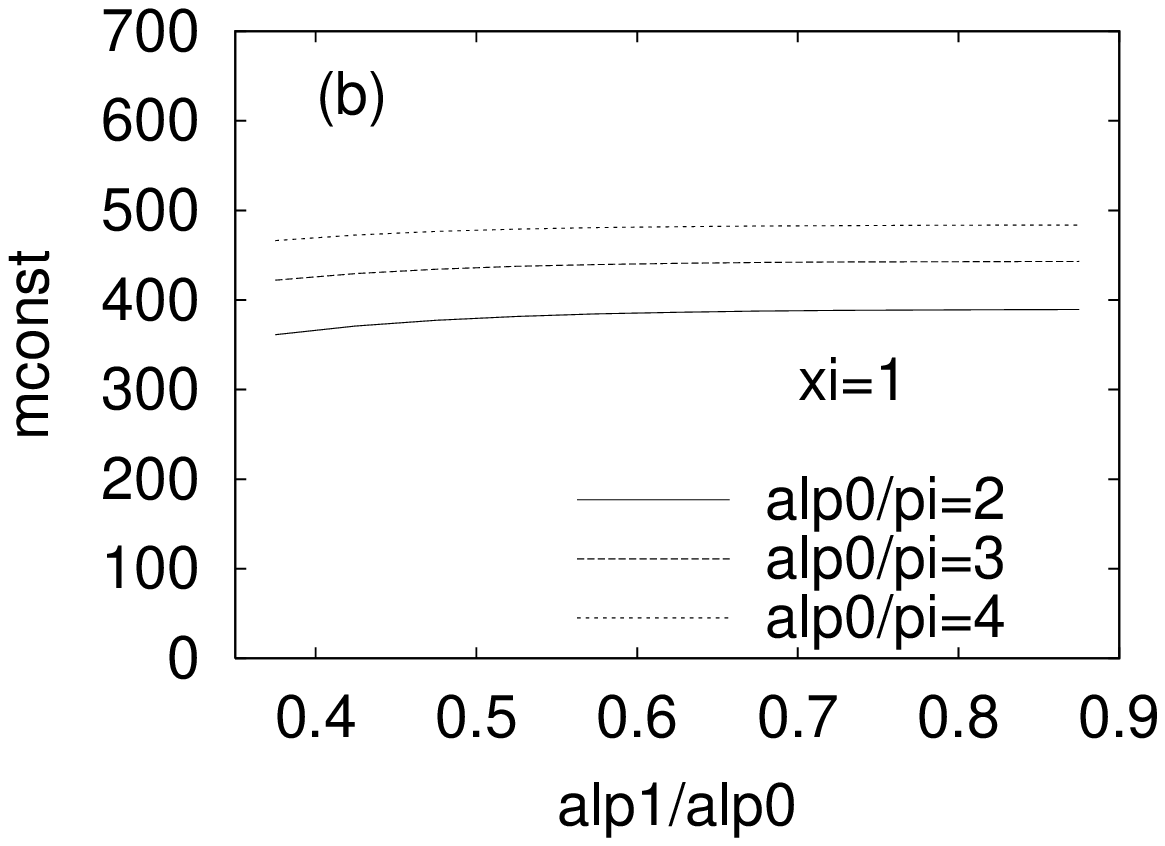}
      \end{center}
    \end{minipage}
  \end{center}
\caption{Results of {\bf (a)} the pion decay constant $f_\pi$  and 
  {\bf (b)} the ``constituent quark mass'' $m_{\rm const}$ with use of the 
  effective coupling (two-loop RGE + finite part) with 
  $\xi_{\rm bg}=1$.
  The non-local gauge fixing method is adopted in the ladder SD equation.
  The boundary condition of RGE is assumed to be 
  $\alpha_s^{\overline{\rm MS}}(m_b)=0.2197$, which  corresponds to
  $\alpha_s^{\overline{\rm MS}}(M_Z)=0.1172$ at the scale of $M_Z$.
  $N_{\rm PS}$ is taken to be $0.8$.
}
\label{fig:gap-eff1}
\end{figure*}

We next turn to the case with the effective coupling, in which the
scale ambiguity is expected to be resolved at the leading-order. 
Since we use the non-local gauge parameter $\xi(z)$
Eq.(\ref{eq:non-local}) in our analysis of
the SD equation, there does not exist a priori choice of the gauge
parameter $\xi_{\rm bg}$ in the effective coupling.
We thus adopt both $\xi_{\rm bg}=0$  (Figure~\ref{fig:gap-eff0}) and 
$\xi_{\rm bg}=1$ (Figure~\ref{fig:gap-eff1}) and compare the results.
Again, the $\alpha_0$-dependence of the results is of non-negligible
order. 
The calculated $f_\pi$ is relatively close to the observed value
$f_\pi=92.4$MeV for $\alpha_0=2$--$4\pi$ with $\xi_{\rm bg}=0$, while
$\xi_{\rm bg}=1$ leads to a little bit larger predictions for the same
range of $\alpha_0$.
Considering that the non-local gauge $\xi(z)$ Eq.(\ref{eq:non-local})
approaches $\xi=0$ in the asymptotic region, $\xi_{\rm bg}=0$ might be
slightly better choice than $\xi_{\rm bg}=1$.

\section{Summary and Discussions}
\label{sec:summary}

In this paper, we have calculated the pion decay constant $f_\pi$ from
the high energy QCD coupling strength 
$\alpha_s^{\overline{\rm MS}}(M_Z)$ by using the improved ladder
Schwinger-Dyson (SD) equation.
The SD equation was analyzed both with and without the
Higashijima-Miransky approximation for its angular integral.
The non-local gauge parameter method was adopted in the analysis
without the Higashijima-Miransky approximation in order to keep the
wave-function factor trivial $A\equiv 1$.
The effective coupling was calculated in the background gauge fixing
method with arbitrary covariant gauge parameter $\xi_{\rm bg}$.
Analyzing the Landau gauge improved ladder SD equation
combined with the $\xi_{\rm bg}=0$ next-to-leading-order effective
coupling, we obtained $f_\pi=85$--$106$MeV depending on the value of
$\alpha_s^{\overline{\rm MS}}(M_Z)=0.1172\pm 0.0020$ within the
Higashijima-Miransky approximation.
Our result impressively agrees with its experimental value
$f_\pi=92.4$MeV and suggests quantitative validity of the improved
ladder SD equation.
It is interesting to compare our result on $f_\pi$ with the previous
analyses~\cite{Aoki:1990eq,Jain:1991pk}
where the value of the calculated $f_\pi$ is less than a half
of its experimental value $f_\pi=92.4$MeV\@. 
In the previous analyses, the renormalization scale $\mu$ was naively
identified with the gluon momentum $\sqrt{|k^2|}$ in the ladder SD
equation. 
Such an identification has the problem of the scale ambiguity,
however, as we pointed out in this paper. 
The improvement achieved in our analysis comes mainly from our
use of the effective coupling, instead of the naive use of the
one-loop $\overline{\rm MS}$ coupling.
The leading-order $\ln\mu$ dependence of 
$\alpha_s^{\overline{\rm MS}}(\mu)$ cancels with the $\mu$-dependence
of the finite correction in our effective coupling.

We found that the effective coupling with $\xi_{\rm bg}=1$ (the
effective coupling derived in the pinch technique (PT)) leads to a
larger value of $f_\pi$. 
This deviation of $f_\pi$ in $\xi_{\rm bg}=1$ may be understood as an
indication of the double counting of the ladder diagrams in the
analysis of the ladder SD equation in $\xi=0$ combined with the PT
effective coupling ($\xi_{\rm bg}=1$).  
In the analysis of the non-local gauge parameter method, 
we were not able to obtain stable results. 
The results, however, were shown to be consistent with the results of the
Higashijima-Miransky approximation in order of magnitude.

We next comment on a different approach to the scale
ambiguity, in which $\alpha_s^{\overline{\rm MS}}(\mu)$ is used with
$\mu$ tuned so as to minimize the finite correction. 
In the case of QED (or QCD in $N_F\rightarrow\infty$ limit), such a
scale is given by $\mu=\sqrt{|k^2|}\exp(-5/6)$ with $k$ being the
momentum of photon (gluon) propagator, while it is 
$\mu=\sqrt{|k^2|}\exp(-205/264)$ in the case of the gluon propagator
($\xi_{\rm bg}=0$) of $N_F=0$ QCD\@.
Since these scales are close to each other, it may be possible to use
universally the scale $\mu=\sqrt{|k^2|}\exp(-5/6)$ in the analysis of
the improved ladder SD equation.
It is straightforward to evaluate the $f_\pi$ in such a simple
prescription.  
Thanks to the scale invariant property of the SD equation, 
we just need to multiply the factor $\exp(5/6)$ to the
result of $f_\pi$ in the naive ladder SD equation with the
$\overline{\rm MS}$ coupling. 
Figure~\ref{fig:hig-maxmin} thus reads $f_\pi=64$--$82$MeV for the 
one-loop $\overline{\rm MS}$ and $f_\pi=122$--$155$MeV for the
two-loop $\overline{\rm MS}$ with this prescription.
The difference between this method and the effective coupling method
comes from the rest of the finite corrections and the treatment of the
two-loop RGE\@. 

Many issues remain unsolved and need further investigation.
In order to calculate the value of $f_\pi$ more precisely with the
non-local gauge parameter method, we need to evaluate the ladder-exact
$f_\pi$ by solving the Bethe-Salpeter equation of the pion.
The problem of the $\xi_{\rm bg}$ dependence of the effective coupling
should also be studied.
It is rather non-trivial task to find adequate relation between $\xi$
(the gauge parameter in the SD equation) and $\xi_{\rm bg}$ (the gauge
parameter in the effective coupling), especially in the non-local
gauge parameter method.
It is also interesting to investigate the critical behavior of the
dense and/or hot QCD using the effective coupling described in this
paper. 
Since the method of the ladder SD equation is now smoothly connected
with the high energy QCD, the high-density and high-temperature
behaviors of the chiral phase transition can now be studied in a more
trustful manner than before.

Finally, the success of the ladder QCD implies that a bulk of driving
force of the dynamical chiral symmetry breaking comes from the
ladder-type diagrams. 
The result presented in this paper therefore provides deeper
understanding of the low energy QCD dynamics.

\begin{acknowledgments}
We thank K.Yamawaki for his useful and helpful comments on an early
draft of this paper. 
We would also like to thank K.Hikasa and Y.Sumino for stimulating
discussions. 
\end{acknowledgments}

\end{document}